\documentclass[aps,pre,twocolumn,superscriptaddress,notitlepage]{revtex4-2}
\usepackage{amsmath}
\usepackage{amsfonts}
\usepackage{amssymb}
\usepackage{lmodern,dsfont}
\usepackage{graphicx}
\usepackage[usenames,dvipsnames]{xcolor}
\usepackage{bm}
\usepackage[english=american]{csquotes}
\usepackage{xr}

\newcommand{\av}[1]{\langle #1 \rangle}
\newcommand{\Av}[1]{\left\langle #1 \right\rangle}
\newcommand{\nn}{\nonumber \\}
\newcommand{\n}{\nonumber}

\newcommand{\grad}{\bm{\nabla}}

\newcommand{\bmc}[1]{\bm{\mathcal{#1}}}

\renewcommand{\eqref}[1]{Eq.~(\ref{#1})}

\begin{document}

\author{Andreas Dechant}
\affiliation{Department of Physics \#1, Graduate School of Science, Kyoto University, Kyoto 606-8502, Japan}
\title{Finite-frequency fluctuation-response inequality}
\date{\today}

\begin{abstract}
We derive an inequality relating the finite-frequency linear response and fluctuations of an observable in a physical system. 
The relation holds for arbitrary observables and perturbations in general Markovian dynamics, including over- and underdamped Langevin systems and jump processes, both in and out of equilibrium.
As a consequence, we obtain a universal upper bound on the broad-band signal-to-noise ratio of noisy dynamics, which only depends on the damping constant and temperature.
We further show that the inequality reduces to an equality for appropriately chosen observables or perturbations in linear systems, both overdamped and underdamped and both in and out of equilibrium.
\end{abstract}

\maketitle

Fluctuations and response are equally important for measurements of small noisy systems:
On the one hand, when probing a system by measuring its response to an external perturbation, the accuracy of the measurement is limited by the fluctuations of the system due to environmental noise.
When the system acts as a sensor, the ratio of response and fluctuations corresponds to the signal-to-noise ratio.
On the other hand, a measurement of the fluctuations themselves can also reveal information about the dynamics of the system.
This connection is most prominent in equilibrium systems, where response and fluctuations are directly proportional and related by the fluctuation-dissipation theorem \cite{Kubo1966,Kubo2012}.

Out of equilibrium, this intimate connection between response and fluctuations is lost.
A phenomenological fluctuation-dissipation theorem may be formulated by introducing an effective temperature \cite{Cugliandolo1997,Loi2008,Cugliandolo2011,Szamel2014,Lippiello2014}, however, the latter depends on the type of measurement \cite{Fielding2002}.
A non-equilibrium fluctuation-dissipation theorem may also be derived from the underlying dynamics \cite{Agarwal1972,Haenggi1982a,Marconi2008,Baiesi2009,Seifert2010a}, expressing the response as fluctuations of the unperturbed system.
However, these fluctuations are expressed using formal conjugate observables that are not directly measurable, thus limiting practical applications.

More recent results establish relations between response and fluctuations even for out-of-equilibrium systems \cite{Dechant2020,Owen2020,FernandesMartins2023,Gao2024,Aslyamov2024,Ptaszynski2024,Liu2025,VanVu2025}, whose common feature is that they are formulated in the form of inequalities rather than equalities like the fluctuation-dissipation theorem.
However, from the point of view of applications, these results suffer from two drawbacks:
First, most results are formulated for discrete systems \cite{Owen2020,FernandesMartins2023,Aslyamov2024,Ptaszynski2024,VanVu2025} and not straightforward to extend to continuous systems.
By contrast, experiments are frequently performed in continuous systems, be it active matter \cite{Loi2011a,Fodor2016}, granular materials \cite{Song2005}, biological systems \cite{Sato2003,Chen2007,Mizuno2008} or levitated nano-particles \cite{Hempston2017,GonzalezBallestero2021}.

Second, and more seriously, existing results only apply to the static response of a system, where the perturbation is constant in time.
Conversely, in experiments, it is more common to work with time-periodic perturbations and to characterize response and fluctuations in frequency space \cite{Lau2003,Brau2007,Wilhelm2008,BenIsaac2011,Loi2011a}.
This has practical reasons, as observing the static response can be challenging in small systems---a tracer particle inside a biological cell subjected to a constant force will eventually collide with the cell wall.
But measuring response at different frequencies also yields additional information about the dynamics and structure of the system that is not contained in the static, zero-frequency response.

In this work, we derive a fluctuation-response inequality (FRI) relating the finite-frequency response and fluctuations of general observables, applicable to both discrete and continuous systems. 
The ratio between the magnitude of the linear response function and the power spectral density of an arbitrary observable at any frequency is upper bounded by a frequency-independent constant, depending on the temperature and friction coefficient of the environment.
As a corollary, we obtain a universal upper bound on the frequency-integrated signal-to-noise ratio of noisy sensors.
Moreover, we show that equality in the FRI is attained for linear dynamics, both in and out of equilibrium, which indicates

\textit{Setup.}
In this work, we focus on Langevin dynamics \cite{Risken1986,Coffey2017} describing the noisy time-evolution of a set of $D$ continuous degrees of freedom $\bm{x}(t) = (x_1(t),\ldots,x_D(t))$.
The dynamics can be overdamped, in which case
\begin{align}
\dot{\bm{x}}(t) = \frac{1}{\gamma} \big(\bm{f}(\bm{x}(t)) + \epsilon \bm{h}(\bm{x}(t),t) \big) + \sqrt{\frac{2 \bm{T}}{\gamma}} \bm{\xi}(t) \label{langevin-over},
\end{align}
or underdamped, in which case
\begin{align}
m \dot{\bm{v}}(t) = - \gamma \bm{v}(t) + \bm{f}(\bm{x}(t)) + \epsilon \bm{h}(\bm{x}(t),t) + \sqrt{2 \gamma \bm{T}} \bm{\xi}(t) \label{langevin-under}.
\end{align}
Here, $\bm{v}(t) = \dot{\bm{x}}(t)$ is the velocity, $\bm{f}(\bm{x})$ are coordinate-dependent forces (interactions or conservative and non-conservative external forces), and $\bm{\xi}(t)$ is a vector of uncorrelated white Gaussian noises. 
All degrees of freedom have the same mass $m$ and friction coefficient $\gamma$ and are in contact with equilibrium environments, whose temperatures are the entries of the diagonal matrix $\bm{T}$.
More general situations, including non-thermal environments, velocity-dependent forces and jump processes on a discrete state space are discussed in Appendix A.
$\bm{h}(\bm{x},t)$ is a perturbation force; we focus on the linear response regime $\epsilon \ll 1$.
We assume that, for $\epsilon = 0$, the dynamics described by \eqref{langevin-over} or \eqref{langevin-under} is in a steady state with probability density $p_\text{st}(\bm{x})$ or $p_\text{st}(\bm{x},\bm{v})$.

We measure a set of $K$ observables $z_k(t) = z_k(\bm{x}(t))$, $k = 1,\ldots,K$, which are functions of the coordinates of the system.
We quantify the fluctuations of the observables using the spectral-density matrix $\bmc{S}(\omega)$, defined as the Fourier transform of the steady-state covariance \cite{Stratonovich1963}
\begin{align}
\mathcal{S}_{kl}(\omega) = \int_{-\infty}^\infty dt \ e^{i \omega t} \text{Cov}(z_k(t),z_l(0)) \label{psd} .
\end{align}
We further consider a set of $Q$ time-dependent perturbations to \eqref{langevin-over} or \eqref{langevin-under},
\begin{align}
\bm{h}(\bm{x},t) = \sum_{q=1}^Q \bm{g}_q(\bm{x}) \phi_q(t) \label{langevin-perturbed} .
\end{align}
Here, $\bm{g}_q(\bm{x})$ are coordinate-dependent forces and $\phi_q(t)$ are time-dependent functions.
In the limit $\epsilon \rightarrow 0$, the response of the observables to the perturbations can to leading order be expressed using the linear-response matrix $\bm{R}(t)$ \cite{Haenggi1982a},
\begin{align}
\lim_{\epsilon \rightarrow 0} \frac{\Av{z_k}_{t,\epsilon} - \Av{z_k}_\text{st}}{\epsilon} = \sum_{q = 1}^{Q} \int_0^t dt' \ R_{k q}(t-t') \phi_q(t') \label{response} ,
\end{align}
where we assume that the system is initially in the steady state with average $\av{z_k}_\text{st}$.
$\Av{z_k}_{t,\epsilon}$ denotes the average value of $z_k(t)$ when applying the perturbation.
$R_{kq}(t-t')$ expresses the impact of the perturbation $q$ at time $t'$ on the average of the observable $k$ at the later time $t$.
Equivalently, we have in Fourier space,
\begin{align}
\bmc{R}(\omega) = \int_0^\infty dt \ e^{i \omega t} \bm{R}(t),
\end{align}
where we used that $\bm{R}(t) = 0$ for $t < 0$.

\textit{Fluctuation-response inequality.}
\begin{figure}
\includegraphics[width=0.49\textwidth]{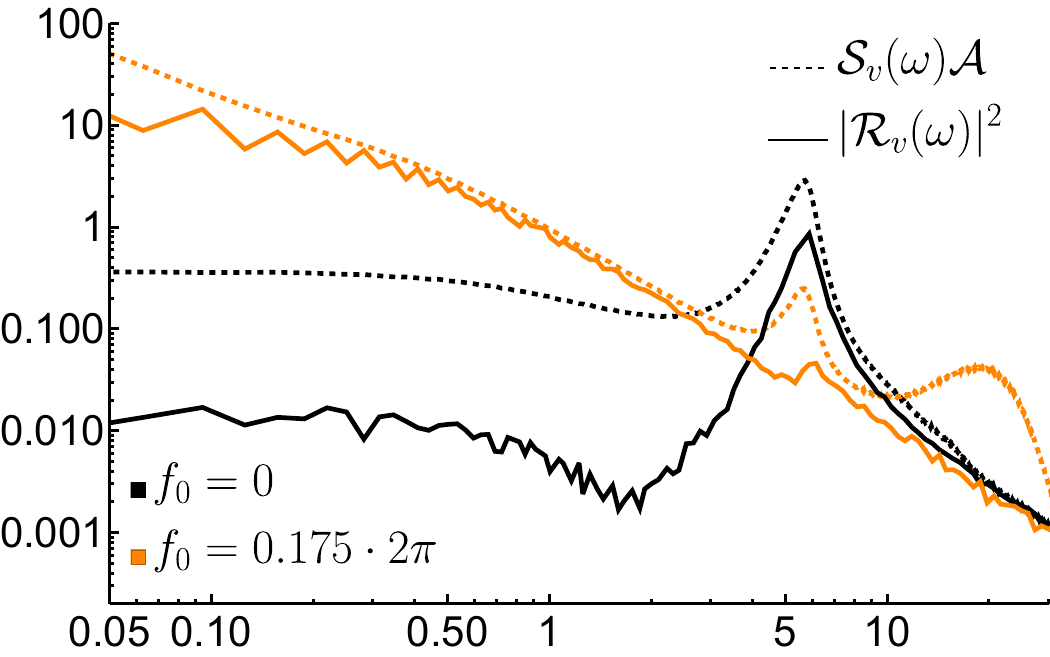}
\caption{Power spectral density (dashed lines) and magnitude of the response (solid lines) of the velocity of an underdampled particle in a periodic potential, \eqref{langevin-perpot}. 
The black lines are for vanishing tilt $f_0 = 0$, where the system is in equilibrium. 
The orange lines are for a non-equilibrium steady state with $f_0 = 0.175 \cdot 2 \pi$. 
Other parameter values are $m = 1, \gamma = 1/3, T = 1/2, U_0 = 1$ and $L = 1$.
Both in and out of equilibrium, the power spectral density upper bounds the response, as predicted by \eqref{fffri-1D}.}
\label{fig-perpot}
\end{figure}
The first main result is the inequality between the response- and spectral-density matrices,
\begin{gather}
\bmc{R}^\text{H}(\omega) \bmc{S}^{-1}(\omega) \bmc{R}(\omega) \leq \bmc{A}  \label{fffri}, \\
\text{with} \qquad \mathcal{A}_{qr} = \frac{1}{2 \gamma} \Av{ \bm{g}_q \cdot \bm{T}^{-1} \bm{g}_r}_\text{st} \n .
\end{gather}
where H denotes the Hermitian conjugate.
The matrix inequality \eqref{fffri} means that the difference of the right-hand and left-hand side is positive semi-definite.
Since the right-hand side is independent of frequency and the choice of the observables, this relation yields a strong constraint between the response and fluctuations of observables in Markovian systems, irrespective of whether the system is in or out of equilibrium.
This generalizes the linear fluctuation-response inequality (FRI) derived in Ref.~\cite{Dechant2020} to more general observables and finite frequencies; we therefore refer to \eqref{fffri} as the finite-frequency FRI.
For a single perturbation $\bm{g}(\bm{x}) \phi(t)$, the response matrix reduces to the vector of response functions and the matrix $\bmc{A}$ to the scalar $\mathcal{A} = \av{\bm{g}\cdot \bm{T} \bm{g}}_\text{st}/(2\gamma)$.
In this case, we definite the response efficiency
\begin{align}
\eta^R(\omega) = \frac{\bmc{R}^*(\omega) \cdot \bmc{S}^{-1}(\omega) \bmc{R}(\omega)}{\mathcal{A}} \leq 1 \label{response-efficiency},
\end{align}
where $*$ denotes complex conjugation.
The response efficiency satisfies $0 \leq \eta^R(\omega) \leq 1$ and quantifies the magnitude of the observed response in relation to the fluctuations of the observables.
It allows characterizing the sensitivity to perturbations in a unified manner, enabling comparisons between different systems, observables and perturbations.
For example, a weakly trapped particle is susceptible to perturbations and exhibits a large response, however, it is also susceptible to fluctuations and its measurement does necessarily not yield more information about the perturbation than a strongly trapped particle with small response and small fluctuations.
For a single observable and a single perturbation ($K = Q = 1$), \eqref{fffri} and \eqref{response-efficiency} simplify to
\begin{align}
\big\vert \mathcal{R}(\omega) \big\vert^2 \leq \mathcal{A} S(\omega), \qquad \eta^R(\omega) = \frac{\vert \mathcal{R}(\omega) \vert^2}{\mathcal{S}(\omega) \mathcal{A}} \label{fffri-1D}.
\end{align}
Thus, the response of any observable to a perturbation with frequency $\omega$ is bounded by the fluctuations of the observable at the same frequency times a constant.
If the functional form of the perturbation is known, then the right hand side can be evaluated from the steady-state statistics of the system.
Therefore, \eqref{fffri-1D} allows us to infer a bound on the response from a passive measurement of the system.

\textit{Response and fluctuations in and out of equilibrium.}
To demonstrate this bound, we consider an underdamped particle in a tilted periodic potential
\begin{align}
m \dot{v}(t) &= - \gamma v(t) - U'(x(t)) + f_0  + \epsilon g_0 \phi(t) + \sqrt{2 \gamma T} \xi(t) \label{langevin-perpot} ,
\end{align}
where $U(x) = U_0 (1-\cos(2 \pi x/L))$.
The constant force $f_0$ tilts the potential and drives the system into a non-equilibrium steady state with a finite average velocity $\av{v}_\text{st}$ at long times.
We remark that, despite the apparent simplicity of \eqref{langevin-perpot}, underdamped diffusion in periodic potentials is not completely understood from a theoretical point of view and possesses a rich structure of interacting dynamical effects \cite{Vollmer1983,Lindenberg2005,Marchenko2012,Lindner2016,Fischer2018,Fischer2020,Spiechowicz2023}.
We focus on the fluctuations and response of the velocity and compute the power spectrum $\mathcal{S}_v(\omega)$ and response $\mathcal{R}_v(\omega)$ from numerical simulations (see Appendix D for details).
The results are shown in Fig.~\ref{fig-perpot}.
In equilibrium, the velocity power spectrum (black dashed line) shows a pronounced peak at $\omega \approx 2 \pi$, corresponding to oscillations of a trapped particle.
Since the particle can escape from the potential by thermal fluctuations, at low frequencies, the velocity spectrum saturates at the long-time diffusion coefficient of the particle,
\begin{align}
D_x = \frac{1}{2} \lim_{\tau \rightarrow \infty} \frac{\text{Var}(x)}{\tau} = \frac{1}{2} \lim_{\omega \rightarrow 0} S_v(\omega) .
\end{align}
At high frequencies, the velocity spectrum decays as $S_v(\omega) \simeq 2 \gamma T/(m^2 \omega^2)$.
The response spectrum (black solid line) likewise exhibits a peak near the oscillation frequency of the particle, whose height, in agreement with \eqref{fffri-1D}, is reduced compared to the one in the power spectrum.
We remark that this effect stems from the non-linearity of the trapping force---for a linear oscillator, \eqref{fffri-1D} turns onto an equality.
Equality in \eqref{fffri-1D} is observed at high frequencies, where fluctuations and response are dominated by the properties of the environment.
Driving the system out of equilibrium by tilting the potential decreases the effective barrier height, allowing the particle to more easily escape from a minimum of the potential.
This reduces the power spectrum at the oscillation frequency, while an additional peak appears at the frequency at which the particle traverses the potential.
Tilting is known to greatly increase the diffusion coefficient \cite{Reimann2001,Lindner2016}, leading to a significant increase in the low-frequency spectrum.
Both the reduction in the power spectrum at the oscillation frequency and the increased low-frequency fluctuations are mirrored in the behavior of the response.
In particular, the value of non-equilibrium power spectrum near the oscillation frequency is lower than the response in equilibrium, and thus \eqref{fffri-1D} predicts a reduction of the non-equilibrium response from a passive measurement of the fluctuations.

\textit{Bound on signal-to-noise ratio.}
The second main result is a consequence of \eqref{fffri-1D} for a single observable $z(t)$ and an equilibrium environment $\bm{T} = T \bm{I}$.
In this case, the relation
\begin{align}
\frac{1}{\pi} \int_0^\infty d\omega \ \mathcal{S}(\omega) = \text{Var}_\text{st}(z)
\end{align}
holds, where $\text{Var}_\text{st}(z)$ denotes the steady-state fluctuations of $z(t)$.
Integrating \eqref{fffri-1D} over frequency yields
\begin{align}
\frac{1}{\pi} \int_0^\infty d\omega \ \big\vert \mathcal{R}(\omega) \big\vert^2 \leq \frac{1}{2 \gamma T} \Av{\Vert \bm{g} \Vert^2}_\text{st} \text{Var}_\text{st}(z) .
\end{align}
The ratio between the magnitude of the response and the fluctuations is commonly referred to as the signal-to-noise ratio (SNR) \cite{Fraden2010},
\begin{align}
\text{SNR}(\omega) = \frac{\big\vert \mathcal{R}(\omega) \big\vert}{\sqrt{\text{Var}_\text{st}(z)}} ,
\end{align}
which allows rewriting the above inequality as
\begin{align}
\int_0^\infty d\omega \ \big(\text{SNR}(\omega)\big)^2 \leq \frac{\pi \av{\Vert \bm{g} \Vert^2}_\text{st}}{2 \gamma T} \label{snr-bound} .
\end{align}
Thus, the frequency-integral of the SNR is bounded by the friction coefficient and temperature of the environment, and the overall magnitude of the perturbation, independent of the precise details of the system.
The left-hand side of \eqref{snr-bound} quantifies the broad-band SNR of the system over all frequencies.
As we show below, equality is attained when the system consists of a single particle in a harmonic confinement.
In that sense, a harmonically trapped particle in equilibrium is the best possible broad-band sensor, achieving the largest possible frequency-integrated SNR.
In practice, however, one is often more interested in the SNR at specific frequencies, which may be substantially enhanced out of equilibrium \cite{Dechant2024}.
In such a case, \eqref{snr-bound} states that enhancing the SNR beyond the single particle value in a given frequency range necessarily leads to a suppression of the SNR at other frequencies.

\textit{Conservation of response in linear networks.}
The third main result is that equality in \eqref{fffri} is attained for linear forces, $\bm{f}(\bm{x}) = - \bm{K}(\bm{x} - \bm{x}_\text{eq})$, where $\bm{K}$ is a matrix and $\bm{x}_\text{eq}$ is the equilibrium position.
In that case, when the observables are the $D$ positions $\bm{x}(t)$ of the particles and the perturbation is a spatially constant force $\bm{g}(\bm{x}) = \bm{g}_0$, we obtain
\begin{align}
\bmc{R}_x^*(\omega) \cdot \bmc{S}_x^{-1}(\omega) \bmc{R}_x(\omega) = \mathcal{A} = \frac{1}{2 \gamma} \bm{g} \cdot \bm{T}^{-1} \bm{g} \label{fffri-linear} .
\end{align}
This relation is shown below for overdamped systems and in Appendix B for underdamped systems.
Importantly, this identity holds irrespective of whether the dynamics is in or out of equilibrium.
For general perturbations, equality only holds when all degrees of freedom of the system are observed.
Moreover, since \eqref{fffri-linear} is an inequality for systems with non-linear forces, it implies that, for the a given magnitude of the response, non-linear systems exhibit larger fluctuations.

\begin{figure}
\includegraphics[width=.35\textwidth]{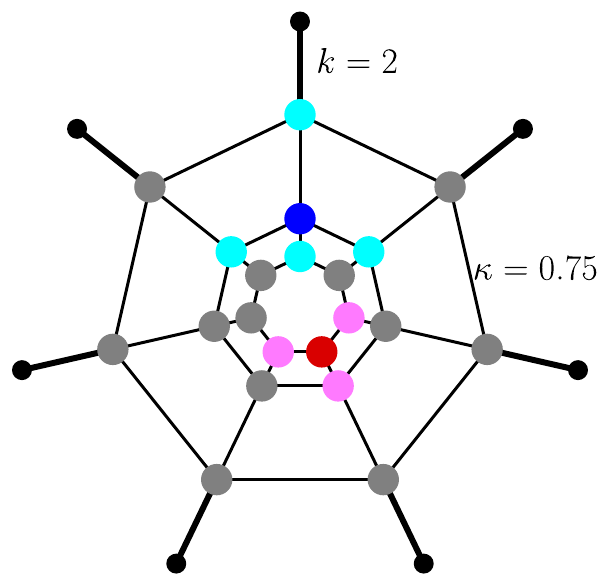}
\caption{A network consisting of 21 oscillators (disks), coupled to each other by springs with spring constant $\kappa = 0.75$; the outermost ring is attached to the black points by springs with spring constant $k = 2$. 
The mass of each oscillator is $m = 1$, the damping constant is $\gamma = 0.01$ and the temperature $T = 1$. 
The perturbation is applied to the red oscillator.}
\label{fig-network}
\end{figure}

\begin{figure}
\includegraphics[width=.49\textwidth]{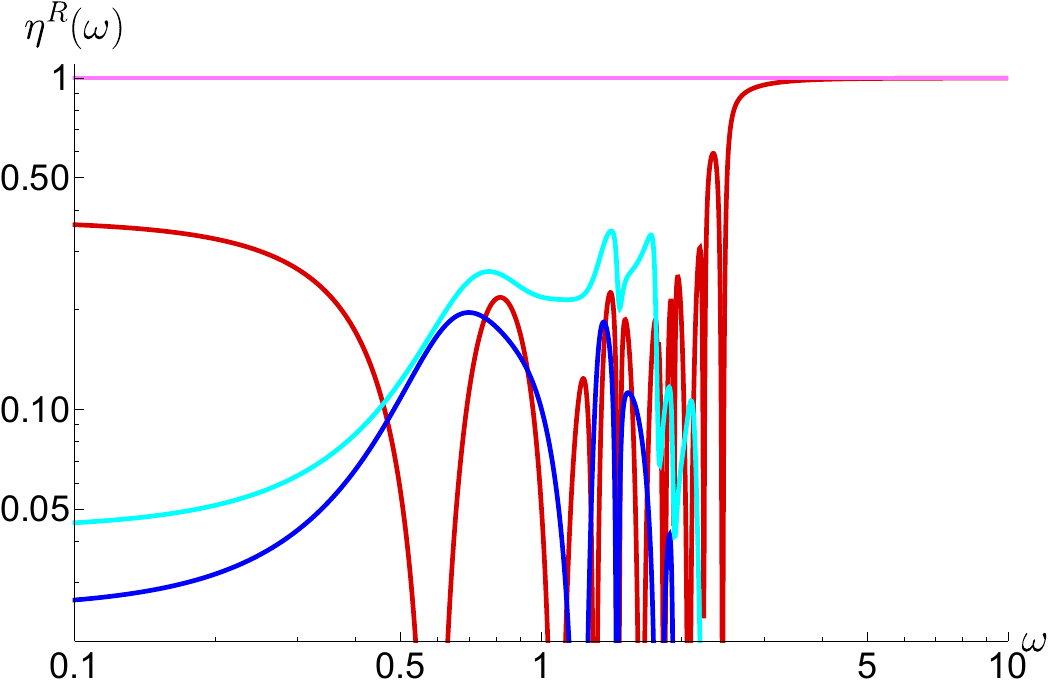}
\caption{Response efficiency of a linear network. 
The response efficiency \eqref{response-efficiency} of the perturbed oscillator (red line) approaches $1$ in the high-frequency limit, where the perturbation is localized. 
At intermediate frequencies, the response efficiency in other parts of the network (blue) can be larger.
Also measuring the response of the neighbors (light blue) yields additional information about the perturbation and increases the response efficiency.
In particular, measuring the perturbed oscillator and its neighbors (light red) results in a response efficiency of unity at all frequencies. }
\label{fig-network-response}
\end{figure}

As an example, we consider a linear network of coupled oscillators, depicted in Fig.~\ref{fig-network}.
When applying a perturbation to one or several oscillators, its effect spreads across the network and we can measure the corresponding response of each oscillator.
\eqref{fffri-linear} states that the overall response of all oscillators, weighted by their fluctuations, is a conserved quantity.
In other words, measuring the response and fluctuations of all oscillators allows recovering the magnitude of an arbitrary perturbation; measuring a subset of oscillators generally only results in a lower bound.
The portion of the perturbation that can be recovered from the measurement is quantified by the response efficiency \eqref{response-efficiency}.
Fig.~\ref{fig-network-response} shows the response efficiency of the perturbed oscillator (red) and another oscillator in the network (blue).
We observe that, in the high-frequency limit, the response efficiency of the perturbed oscillator approaches unity; at high frequencies, response and fluctuations are determined by the local properties of the environment.
At intermediate frequencies, the perturbation excites collective modes in the network and causes a non-vanishing response of other oscillators.
Interestingly, the response efficiency can be larger for oscillators other than the perturbed one, indicating that a non-local measurement can yield more information about the perturbation.

Measuring more than one oscillator simultaneously provides additional information about the perturbation.
Measuring the neighboring (light blue) oscillators in addition to the blue one results in increased response efficiency (light blue line in Fig.~\ref{fig-network-response}).
Strikingly, when measuring the perturbed (red) oscillator and its neighbors (light red) simultaneously, we find an response efficiency of unity for all frequencies (light red line in Fig.~\ref{fig-network-response}).
As we prove in Appendix C, this finding is generic: In linear networks, the magnitude of the perturbation can always be inferred from a measurement of the perturbed oscillators and their neighbors.

\textit{Proof of the FRI for overdamped dynamics.}
In the interest of conciseness, we focus on the overdamped case \eqref{langevin-over} here; the derivation for more general overdamped dynamics, as well as underdamped Langevin and Markov jump dynamics is provided in SM A.
We consider a set of $K$ observables of the form
\begin{align}
r_k(t) = z_k(\bm{x}(t)) + \bm{w}_k(\bm{x}(t)) \circ \dot{\bm{x}}(t) \label{observable-current} .
\end{align}
This is slightly more general than the form discussed above, since it also contains a current-like contribution, where $\circ$ denotes the Stratonovich product.
We denote the steady-state average of $r_k(t)$ by $\Av{r_k}_\text{st}$.
Next, the response of the observables is calculated.
A central result of non-equilibrium linear response theory \cite{Agarwal1972,Haenggi1982a,Marconi2008,Baiesi2009,Seifert2010a} is that the response function can be formally expressed as a steady-state correlation function
\begin{align}
R_{k q}(t) = \Av{r_k(t) \psi_q(0)} .
\end{align}
The first crucial insight is that the quantity $\psi_q(t)$ has the same form as the observables \eqref{observable-current}; specifically,
\begin{align}
\psi_q(t) &= \tilde{z}_q(\bm{x}(t)) + \tilde{\bm{w}}_q(\bm{x}(t)) \circ \dot{\bm{x}}(t) \\
\tilde{z}_q(\bm{x}) &= - \frac{1}{2 \gamma^2} \bm{g}_q(\bm{x}) \cdot \bm{T}^{-1} \bm{f}(\bm{x}) - \frac{1}{2 \gamma} \grad_x \cdot \bm{g}_q(\bm{x}) \nn
\tilde{\bm{w}}_q(\bm{x}) &= -\frac{1}{2 \gamma^2} \bm{T}^{-1} \bm{g}_q(\bm{x}) \n .
\end{align}
The second crucial insight is that the correlation function of the \enquote{response observables} $\psi_q(t)$ is given by
\begin{align}
\Av{\psi_q(t) \psi_r(t')} = \frac{1}{2 \gamma} \Av{\bm{g}_q \cdot \bm{T}^{-1} \bm{g}_r}_\text{st} \delta(t-t') ,
\end{align}
which can be obtained by direct calculation, see Appendix A.
This result allows writing the joint spectral density matrix $\mathbb{S}(\omega)$ of the $K+Q$ observables $(r_1(t),\ldots,r_k(t),\psi_1(t),\ldots,\psi_Q(t))$ in block-wise form
\begin{align}
\mathbb{S}(\omega) = \begin{pmatrix} \bmc{S}(\omega) & \bmc{R}(\omega) \\ \bmc{R}^\text{H}(\omega) & \bmc{A} \end{pmatrix},
\end{align}
where the symmetric, positive definite matrix $\bmc{A}$ is defined as
\begin{align}
\mathcal{A}_{q r} = \int_{-\infty}^\infty dt \ e^{i \omega t} \Av{\psi_q(t) \psi_r(0)} = \frac{1}{2 \gamma} \Av{\bm{g}_q \cdot \bm{T}^{-1} \bm{g}_r}_\text{st} .
\end{align}
Since $\mathbb{S}(\omega)$ is a positive definite matrix, its blocks have to satisfy the conditions
\begin{subequations}
\begin{gather}
\bmc{R}^\text{H}(\omega) \bmc{S}^{-1}(\omega) \bmc{R}(\omega) \leq \bmc{A}, \label{response-tradeoff-1} \\
\bmc{R}(\omega) \bmc{A}^{-1} \bmc{R}^\text{H}(\omega) \leq \bmc{S}(\omega) \label{response-tradeoff-2} .
\end{gather} \label{response-tradeoff}%
\end{subequations}
The first condition is precisely the first main result, \eqref{fffri}.
The second condition provides an alternative trade-off relation between the fluctuations and the response.
Integrating the second condition with respect to frequency yields, for configuration-dependent observables with $\bm{w}_q = 0$,
\begin{align}
\frac{1}{\pi} \int_{0}^\infty d\omega \ \bmc{R}(\omega) \bmc{A}^{-1} \bmc{R}^\text{H}(\omega) \leq \bm{\Xi}_\text{st}.
\end{align}
That is, the frequency-integral of the squared response matrix is bounded by the steady-state covariance matrix of the observables.
This generalizes \eqref{snr-bound} to the case of several observables and perturbations.

\textit{Linear dynamics.}
Next, the above results are applied to a system of particles with linear interactions and perturbations that are constant in space.
In this case, \eqref{langevin-perturbed} is written as
\begin{align}
\dot{\bm{x}}(t) = \frac{1}{\gamma} \Big(-\bm{K} \big(\bm{x}(t) - \bm{x}_\text{eq} \big) + \epsilon \bmc{G} \bm{\phi}(t) \Big)  + \sqrt{\frac{2 \bm{T}}{\gamma}} \bm{\xi}(t) \label{langevin-linear} .
\end{align}
The matrix $\bmc{G} \in \mathbb{R}^{D \times Q}$ has the constant force $\bm{g}_q$ as its $q$-th column and $\bm{\phi}(t) = (\phi_1(t),\ldots,\phi_Q(t))$.
$\bm{K}$ is a matrix of spring constants, whose eigenvalues have positive real parts, so that the system has a stable steady state.
If $\bm{K}$ is symmetric, the resulting steady state is in equilibrium; if $\bm{K}$ is not symmetric, then non-reciprocal interactions drive the system into a non-equilibrium steady state.
Since both the dynamics and the perturbation in \eqref{langevin-linear} are linear, the linear response matrix \eqref{response} fully describes the response of the system for arbitrary magnitude $\epsilon$ of the perturbation.
The observables are likewise linear combinations of the coordinates of the particles, i.~e.~$\bm{z}(t) = \bmc{Z} \bm{x}(t)$ with $\bmc{Z} \in \mathbb{R}^{K \times D}$.
Since the system is linear, explicit expressions for the spectral density and response matrices are obtained (see Appendix B),
\begin{align}
\bmc{S}(\omega) &= \gamma \bmc{Z} \Big( \big( \bm{K} - i \omega \gamma \bm{I} \big)^{-1} \bm{\Xi}_{\text{st}} + [\text{h.c.}] \Big) \bmc{Z}^\text{T}, \nn
\bmc{R}(\omega) &= \bmc{Z} \big( \bm{K} - i\omega \gamma \bm{I} \big)^{-1} \bmc{G} , \\
\bmc{A} &= \frac{1}{2 \gamma} \bmc{G}^\text{T} \bm{T}^{-1} \bmc{G} \n .
\end{align}
Here, [h.c.] represents the Hermitian conjugate of the first matrix and $\bm{\Xi}_\text{st}$ denotes the steady-state covariance matrix of the coordinates.
With these expressions, it is a matter of straightforward linear algebra (see Appendix B) to verify the relations
\begin{subequations}
\begin{align}
\bmc{R}^\text{H}(\omega) \bmc{S}^{-1}(\omega) \bmc{R}(\omega) = \bmc{A} \quad &\text{if} \; \text{rank}(\bmc{Z}) = D \label{linear-identity-1} \\
\bmc{R}(\omega) \bmc{A}^{-1} \bmc{R}^\text{H}(\omega) = \bmc{S}(\omega) \quad &\text{if} \; \text{rank}(\bmc{G}) = D \label{linear-identity-2}.
\end{align}\label{linear-identity}%
\end{subequations}
If the observables provide the complete information about the coordinates of the particles ($\bmc{Z}$ is an invertible matrix), then \eqref{response-tradeoff-1} is an equality.
Conversely, if the perturbations can reproduce any constant-in-space force ($\bmc{G}$ is an invertible matrix), then \eqref{response-tradeoff-2} is an equality.
Equality holds for both equilibrium and out-of-equilibrium situations, as long as the dynamics is linear.

\textit{Discussion.}
In this work, we derived a universal trade-off relation between the frequency response and the frequency-resolved fluctuations of discrete and continuous Markovian dynamics.
This fluctuation-response inequality quantifies the intuition that response and fluctuations are related even out of equilibrium, since they are both properties of the same underlying physical system.
We stress that \eqref{fffri} does not reduce to an equality in equilibrium and is thus distinct from the fluctuation-dissipation theorem.
Instead, we found that equality is attained for linear dynamics, independent of whether they are in or out of equilibrium.
Whether and how \eqref{fffri} can be related to the fluctuation-dissipation theorem and to the computation of dissipation from its violation \cite{Harada2005,Harada2006} is an open question.

The derivation of \eqref{fffri} only requires the response to be represented as a correlation function of the unperturbed system.
Since this relation holds for general Markovian dynamics and possibly beyond, we expect that the result can be readily generalized to other dynamics, such as open quantum systems, where the corresponding zero-frequency result was recently obtained in Ref.~\cite{VanVu2025}.
A more challenging generalization is to derive similar bounds on non-linear response.
Recently, a relation between static non-linear response and first passage times was derived in Ref.~\cite{Bao2025}, but whether a similar result also applies to time-periodic perturbations is unclear.
Another approach is to characterize the system in the presence of a finite-strength periodic perturbation using Floquet theory and then apply the non-linear FRI derived in Ref.~\cite{Dechant2020}.

The finite-frequency FRI \eqref{fffri} does not distinguish between systems that are in or out of equilibrium.
However, the relation between response and fluctuations is clearly different out of equilibrium, where the fluctuation-dissipation theorem does not hold.
Therefore, an important question is whether there are bounds on the response that explicitly take into account the non-equilibrium nature of the system.
One such relation was derived for the static response of currents in Ref.~\cite{Ptaszynski2024}, which is bounded by the entropy production rate.
Similar relations for the finite-frequency response might be obtained starting from the variational formula for the power-spectral density derived in Refs.~\cite{Dechant2023a,Vo2025}, which provides a formal identity between response the power spectrum, from which bounds can be derived.

\widetext

\appendix

\section{Generalized fluctuation-dissipation theorem and fluctuation-response inequality in frequency space} \label{sec-fffri}


\subsection{Overdamped Langevin dynamics} \label{sec-fffri-over}
We consider the overdamped Ito-Langevin equations for $\bm{x} \in \mathbb{R}^D$,
\begin{align}
\dot{\bm{x}}(t) = \bm{a}(\bm{x}(t)) + \epsilon \sum_{q = 1}^{Q} \bm{y}_q(\bm{x}(t)) \phi_q(t) + \sqrt{2 \bm{B}(\bm{x}(t))} \cdot \bm{\xi}(t) \label{langevin} .
\end{align}
Here $\bm{a}(\bm{x})$ is the drift vector and $\bm{B}(\bm{x})$ is the positive definite diffusion matrix; $\sqrt{\bm{B}(\bm{x})}$ denotes its unique positive definite matrix square root.
$\odot$ denotes the combination of matrix- and Ito-product, and $\bm{\xi}(t)$ is a vector of mutually uncorrelated, Gaussian white noises, $\Av{\xi_h(t) \xi_j(t')} = \delta_{h j} \delta(t-t')$.
The vectors $\bm{y}_q(\bm{x})$ represent small perturbations ($\epsilon \ll 1$) to the drift vector, whose time-dependence is encoded by the functions $\phi_q(t)$.
\eqref{langevin-over} of the main text is recovered for $\bm{a}(\bm{x}) = \frac{1}{\gamma} \bm{f}(\bm{x})$, $\bm{y}_q(\bm{x}) = \frac{1}{\gamma} \bm{g}_q(\bm{x})$ and $\bm{B}(\bm{x}) = T/\gamma \bm{I}$ with $\bm{I}$ the identity matrix.
We assume that, without the perturbation, the system is in a steady state characterized by the probability density $p_\text{st}(\bm{x})$, which obeys the steady-state Fokker-Planck equation \cite{Risken1986}
\begin{align}
0 &= - \grad_x \big( \bm{\nu}_\text{st}(\bm{x}) p_\text{st}(\bm{x}) \big) \quad \text{with} \label{fpe-steady} \\
\bm{\nu}_\text{st}(\bm{x}) &= \bm{a}(\bm{x}) - \grad \bm{B}(\bm{x}) - \bm{B}(\bm{x}) \grad \ln p_\text{st}(\bm{x}) \n ,
\end{align}
where $\bm{\nu}_\text{st}(\bm{x})$ is called the steady-state local mean velocity \cite{Seifert2010a}, which quantifies the magnitude of flows in the (generally) non-equilibrium steady state.
Here, we defined the vector-valued divergence of a matrix by $(\grad \bm{B}(\bm{x}))_{j} = \sum_{h} \partial_{x_h} B_{h j}(\bm{x})$.
In order to avoid confusion with the Ito-product, we omit the scalar product in denoting the divergence; any product between two vector-valued quantities is implicitly assumed to be a scalar product unless noted otherwise.
Note that the steady state is in equilibrium (satisfied detailed balance) if and only if the local mean velocity vanishes identically; the steady-state entropy production rate is
\begin{align}
\sigma_\text{st} = \Av{\bm{\nu}_\text{st} \bm{B}^{-1} \bm{\nu}_\text{st}}_\text{st},
\end{align}
which is strictly positive for any non-zero $\bm{\nu}_\text{st}(\bm{x})$ as $\bm{B}(\bm{x})$ is assumed to be positive definite.

\subsubsection{Observables and power-spectral density}
We focus on $K$ scalar observables that each consist of a configuration-dependent contribution $z_k(\bm{x}(t))$ and a current-like contribution $\bm{w}_k(\bm{x}(t)) \circ \dot{\bm{x}}(t)$, where $\circ$ is the Stratonovich-product,
\begin{align}
r_k(t) = z_k(\bm{x}(t)) + \bm{w}_k(\bm{x}(t)) \circ \dot{\bm{x}}(t) \label{observable} .
\end{align}
The steady-state average of $r_k(t)$ is given by
\begin{align}
\Av{r_k}_\text{st} = \int d\bm{x} \ \Big( z_k(\bm{x}) + \bm{w}_k(\bm{x}) \bm{\nu}_\text{st}(\bm{x}) \Big) p_\text{st}(\bm{x}) =  \Av{z_k}_\text{st} + \Av{\bm{w}_k \bm{\nu}_\text{st}}_\text{st} .
\end{align}
We characterize the fluctuations of the observables in the steady state using their covariances,
\begin{align}
C_{kl}(t) = \text{Cov}_\text{st}(r_k(t),r_l(0)) = \Av{r_k(t) r_l(0)}_\text{st} - \Av{r_k}_\text{st} \Av{r_l}_\text{st} ,
\end{align}
where the subscript st indicates that the average is taken with respect to the unperturbed steady state for $\epsilon = 0$.
Due to time-translation invariance, the covariance matrix $\bm{C}(t)$ only depends on the time-lag $t$.
However, since the system described by \eqref{langevin} is generally out of equilibrium even in the steady state, $\bm{C}(t)$ is not a symmetric matrix; rather we have $\bm{C}^\text{T}(t) = \bm{C}(-t)$.
Equivalently, we can characterize the correlations by the power-spectral density matrix, which is the Fourier-transform of the covariance matrix $\bm{C}(t)$,
\begin{align}
\bmc{S}(\omega) = \int_{-\infty}^\infty dt \ e^{i \omega t} \bm{C}(t) \label{power-spectral-density} .
\end{align}
The power-spectral density $\bmc{S}(\omega)$ is a Hermitian, positive definite matrix.
It is real and symmetric when $\bm{C}(t)$ is symmetric, that is, when the system \eqref{langevin} is in equilibrium and thus time-reversal symmetric.
Note that the power spectrum can equivalently be defined by introducing the finite-time Fourier transform of the observables,
\begin{align}
\hat{r}_k(\omega,\tau) = \int_0^\tau dt \ e^{i \omega t} r_k(t).
\end{align}
By the Wiener-Khinchine theorem, we then obtain
\begin{align}
\mathcal{S}_{kl}(\omega) = \lim_{\tau \rightarrow \infty} \frac{1}{\tau} \text{Cov}_\text{st}\big(\hat{r}_k(\omega,\tau),\hat{r}_l^*(\omega,\tau) \big),
\end{align}
where $*$ denotes complex conjugation.
This identity shows that $\bmc{S}(\omega)$ is itself a covariance matrix of complex observables, which implies its positive-definiteness.

\subsubsection{Response and fluctuation-dissipation theorem}
Next, we want to characterize the response of the observables to the perturbations.
The change in the average value of an observable is
\begin{align}
\Delta \Av{r_k(t)} = \Av{r_k(t)}_\epsilon - \Av{r_k}_\text{st} ,
\end{align}
where $\Av{r_k}_\epsilon$ denotes the average of the observable for $\epsilon > 0$.
We focus on the linear response regime $\epsilon \rightarrow 0$, where we have
\begin{align}
\lim_{\epsilon \rightarrow 0} \frac{\Delta \Av{r_k(t)}}{\epsilon} = \partial_\epsilon \Av{r_k(t)}_\epsilon \Big\vert_{\epsilon = 0} .
\end{align}
Suppose that the perturbation vanishes, that is $\phi_q(t) = 0$ for all $q$, for $t < 0$.
Then, we can formally write
\begin{align}
\Av{r_k(t)}_\epsilon = \int d\Gamma \ r_k(t) \mathbb{P}_\epsilon(\Gamma) .
\end{align}
Here, $\Gamma = \lbrace\bm{x}(t')\rbrace_{t' \in [0,t]}$ is a trajectory of the system during the time-interval $[0,t]$ and $\mathbb{P}_\epsilon(\Gamma)$ is the probability of the trajectory in the perturbed system, starting from the steady state at $t = 0$.
Since the trajectory probability is positive, we can write
\begin{align}
\partial_\epsilon \Av{r_k(t)}_\epsilon \Big\vert_{\epsilon = 0} &= \int d\Gamma \ r_k(t) \partial_\epsilon \mathbb{P}_\epsilon(\Gamma) \Big\vert_{\epsilon = 0} = \int d\Gamma \ r_k(t) \partial_\epsilon \big(\ln \mathbb{P}_\epsilon(\Gamma) \big) \mathbb{P}_\epsilon(\Gamma) \Big\vert_{\epsilon = 0} \nn
& = \int d\Gamma \ r_k(t) \partial_\epsilon \big(\ln \mathbb{P}_\epsilon(\Gamma) \big) \Big\vert_{\epsilon = 0} \mathbb{P}_0(\Gamma) .
\end{align}
We define the trajectory-dependent quantity
\begin{align}
\partial_\epsilon \big(\ln \mathbb{P}_\epsilon(\Gamma) \big) \Big\vert_{\epsilon = 0} = Q(\Gamma) .
\end{align}
Since $\mathbb{P}_0(\Gamma)$ is the trajectory probability of the unperturbed steady-state system, we can write
\begin{align}
\partial_\epsilon \Av{r_k(t)}_\epsilon \Big\vert_{\epsilon = 0} = \Av{r_k(t) Q(\Gamma)}_\text{st} .
\end{align}
As shown in Ref.~\cite{Dechant2020}, the derivative of the logarithm of the path probability is given by
\begin{align}
\partial_\epsilon \big(\ln \mathbb{P}_\epsilon(\Gamma) \big) \Big\vert_{\epsilon = 0} = \frac{1}{2} \int_0^t dt' \ \sum_{q = 1}^Q \phi_q(t') \bm{y}_q(\bm{x}(t')) \bm{B}(\bm{x}(t'))^{-1} \sqrt{2 \bm{B}(\bm{x}(t'))} \cdot \bm{\xi}(t') \equiv \int_0^t dt' \ \sum_{q=1}^Q \psi_q(t') \phi_q(t'), \label{path-log}
\end{align}
which defines the \enquote{response observable}
\begin{align}
\psi_q(t) = \frac{1}{2} \bm{y}_q(\bm{x}(t)) \bm{B}(\bm{x}(t))^{-1} \sqrt{2 \bm{B}(\bm{x}(t))} \cdot \bm{\xi}(t) \label{response-observable} .
\end{align}
This allows us to express the linear response as
\begin{align}
\partial_\epsilon \Av{r_k(t)}_\epsilon \Big\vert_{\epsilon = 0} = \sum_{q = 1}^Q \int_0^t dt' \ \Av{r_k(t) \psi_q(t')}_\text{st} \phi_q(t') .
\end{align}
Since the steady state is time-translation invariant, we can define the response matrix
\begin{align}
R_{kq}(t-t') = \Av{r_k(t) \psi_q(t')}_\text{st} \label{response-correlation},
\end{align}
which results in the expression for the response
\begin{align}
\partial_\epsilon \Av{r_k(t)}_\epsilon \Big\vert_{\epsilon = 0} = \int_0^t dt' \ \sum_{q = 1}^Q R_{kq}(t-t') \phi_q(t') .
\end{align}
Thus, the response matrix quantifies the response of observable $k$ at time $t$ to perturbation $q$ at the earlier time $t'$.
\eqref{response-correlation} is the generalized fluctuation-dissipation theorem discussed in Refs.~\cite{Agarwal1972,Haenggi1982a,Marconi2008,Baiesi2009,Seifert2010a}.
It expresses the response in terms of a correlation function in the unperturbed steady state, at the expense that it involves the correlation of the observable $r_k(t)$ with the quantity $\psi_q(t)$, which is generally not directly measurable.
Since $\psi_q(t)$ involves an Ito-product with the noise, we obtain $R_{kq}(t-t') = 0$ for $t' > t$ due to the non-anticipating property of white noise, i.~e.~the noise at time $t'$ is independent of the dynamics at any earlier time $t$.
Using that, in the unperturbed system, we have from \eqref{langevin}
\begin{align}
\sqrt{2 \bm{B}(\bm{x}(t))} \cdot \bm{\xi}(t) = \dot{\bm{x}}(t) - \bm{a}(\bm{x}(t)),
\end{align}
we can further write the response observable as
\begin{align}
\psi_q(t) = \frac{1}{2} \bm{y}_q(\bm{x}(t)) \bm{B}(\bm{x}(t))^{-1} \cdot \big(\dot{\bm{x}}(t) - \bm{a}(\bm{x}(t))\big) .
\end{align}
Using Ito's lemma to transform the Ito-product into a Stratonovich product, this can be written as
\begin{align}
\psi_q(t) = - \frac{1}{2} \bm{y}_q(\bm{x}(t)) \bm{B}(\bm{x})^{-1} \bm{a}(\bm{x}(t)) + \frac{1}{2} \bm{y}_q(\bm{x}(t)) \bm{B}^{-1}(\bm{x}) \circ \dot{\bm{x}}(t) - \frac{1}{2} \text{tr} \Big( \bm{B}(\bm{x}(t)) \bm{J}_{\bm{B}^{-1} \bm{y}_q}(\bm{x}(t)) \Big) ,
\end{align}
where tr denotes the trace and $\bm{J}_{\bm{B}^{-1} \bm{y}_q}(\bm{x})$ is the Jacobian matrix of the vector field $\bm{B}(\bm{x})^{-1} \bm{y}_q(\bm{x})$.
Writing the last term explicitly in index notation, he have
\begin{align}
\text{tr} \Big( \bm{B}(\bm{x}) \bm{J}_{\bm{B}^{-1} \bm{y}_q}(\bm{x}) \Big)  = B_{jk}(\bm{x}) \partial_{x_j} \Big( \big(\bm{B}(\bm{x})^{-1} \big)_{k l} y_{q,l}(\bm{x}) \Big) ,
\end{align}
where the sum over repeated indices is implied.
Using the formula for the derivative of an inverse matrix,
\begin{align}
\partial_{x_j} \bm{B}(\bm{x})^{-1} = - \bm{B}(\bm{x})^{-1} \partial_{x_j} \bm{B}(\bm{x}) \bm{B}(\bm{x})^{-1},
\end{align}
we obtain
\begin{align}
\text{tr} \Big( \bm{B}(\bm{x}) \bm{J}_{\bm{B}^{-1} \bm{y}_q}(\bm{x}) \Big)  = B_{jk}(\bm{x}) \big(\bm{B}(\bm{x})^{-1} \big)_{kl} \partial_{x_j} y_{q,l} - B_{jk}(\bm{x}) \big(\bm{B}(\bm{x})^{-1} \big)_{km} \partial_{x_j} B_{mn}(\bm{x}) \big(\bm{B}(\bm{x})^{-1} \big)_{nl} y_{q,l}(\bm{x}) .
\end{align}
Canceling the matrix $\bm{B}(\bm{x})$ and its inverse, this simplifies to
\begin{align}
\text{tr} \Big( \bm{B}(\bm{x}) \bm{J}_{\bm{B}^{-1} \bm{y}_q}(\bm{x}) \Big) = \grad \bm{y}_q(\bm{x}) - \grad \bm{B}(\bm{x}) \bm{B}(\bm{x})^{-1} \bm{y}_q(\bm{x}) .
\end{align}
Recalling the definition of the steady-state local mean velocity \eqref{fpe-steady}, we obtain
\begin{align}
\psi_q(t) = - \frac{1}{2} \bigg( \bm{y}_q(\bm{x}(t)) \bm{B}(\bm{x}(t))^{-1} \bm{\nu}_\text{st}(\bm{x}(t)) + \frac{\grad \big(\bm{y}_q(\bm{x}(t)) p_\text{st}(\bm{x}(t)) \big)}{p_\text{st}(\bm{x}(t))} \bigg) + \frac{1}{2} \bm{y}_q(\bm{x}(t)) \bm{B}(\bm{x}(t))^{-1} \circ \dot{\bm{x}}(t) .
\end{align}
We see that this is of the same form as the observables $r_k(t)$ defined in \eqref{observable},
\begin{gather}
\psi_q(t) = \tilde{z}_q(\bm{x}(t)) + \tilde{\bm{w}}_q(\bm{x}(t)) \circ \dot{\bm{x}}(t) \qquad \text{with} \label{response-observable-2} \\
\tilde{z}_q(\bm{x}) = - \frac{1}{2} \bigg( \bm{y}_q(\bm{x}) \bm{B}(\bm{x})^{-1} \bm{\nu}_\text{st}(\bm{x}) + \frac{\grad \big(\bm{y}_q(\bm{x}) p_\text{st}(\bm{x}) \big)}{p_\text{st}(\bm{x})} \bigg) \qquad \text{and} \qquad \tilde{\bm{w}}_q(\bm{x}) = \frac{1}{2} \bm{B}(\bm{x})^{-1} \bm{y}_q(\bm{x}) . \n
\end{gather}
Thus, the \enquote{response observable}, while not directly measurable, as it requires knowledge about the diffusion matrix, steady-state probability density, and local mean velocity, is mathematically an observable of the same type as $r_k(t)$.

\subsubsection{Fluctuation-response inequality}
We define the frequency-response matrix as the Fourier transform of the response matrix \eqref{response-correlation},
\begin{align}
\bmc{R}(\omega) = \int_0^\infty dt \ e^{i \omega t} \bm{R}(t),
\end{align}
where we used that the response matrix vanishes for $t < 0$.
Since, from \eqref{response-observable-2}, the quantities $\psi_q(t)$ are of the same type as $r_k(t)$, we can consider the power-spectral density matrix of the extended set ob observables $\lbrace r_1(t), \ldots, r_K(t), \psi_1(t), \ldots, \psi_Q(t) \rbrace$,
\begin{align}
\big(\mathbb{S}(\omega) \big)_{k,l} = \int_{-\infty}^\infty dt \ e^{i \omega t} \times \left\lbrace \begin{array}{ll}
\text{Cov}_\text{st}(r_k(t),z_l(0)) &\text{if} \; k,l \leq K \\[1ex]
\text{Cov}_\text{st}(r_k(t),\psi_l(0)) &\text{if} \; k \leq K, l \geq K+1 \\[1ex]
\text{Cov}_\text{st}(\psi_k(t),r_l(0)) &\text{if} \; l \leq K, k \geq K+1 \\[1ex]
\text{Cov}_\text{st}(\psi_k(t),\psi_l(0)) &\text{if} \; k,l \geq K+1 .
\end{array} \right.
\end{align}
From \eqref{response-correlation}, we identify the covariance between $r_k(t)$ and $\psi_q(0)$ as the $(k,q)$ entry of the response matrix and can thus write
\begin{align}
\mathbb{S}(\omega) = \begin{pmatrix} \bmc{S}(\omega) & \bmc{R}(\omega) \\ \bmc{R}(\omega)^\text{H} & \bm{\Psi}(\omega) \end{pmatrix} ,
\end{align}
where $\bm{\Psi}(\omega)$ is the Fourier transform of the covariance of the observables $\psi_q(t)$ and H denotes the Hermitian conjugate.
Note that $\Av{\psi_q}_\text{st} = 0$, which is apparent from \eqref{response-observable}.
From the latter, we also find
\begin{align}
\text{Cov}_\text{st}(\psi_q(t),\psi_r(0)) &= \Av{\psi_q(t) \psi_r(0)}_\text{st} \label{response-observable-correlation} \\ 
&= \frac{1}{4} \Av{\Big(\bm{y}_q(\bm{x}(t)) \bm{B}(\bm{x}(t))^{-1} \sqrt{2 \bm{B}(\bm{x}(t))} \cdot \bm{\xi}(t) \Big) \Big( \bm{y}_r(\bm{x}(0)) \bm{B}(\bm{x}(0))^{-1} \sqrt{2 \bm{B}(\bm{x}(0))} \cdot \bm{\xi}(0) \Big)}_\text{st}\nn
& = \frac{1}{2} \Av{\bm{y}_q \bm{B}^{-1} \bm{y}_r}_\text{st} \delta(t) \n,
\end{align}
which follows from the delta-correlated white noise.
Thus, the matrix $\bm{\Psi}(\omega)$ is independent of frequency,
\begin{align}
\Psi_{qr}(\omega) = \frac{1}{2} \Av{\bm{y}_q \bm{B}^{-1} \bm{y}_r}_\text{st}  = \mathcal{A}_{qr},
\end{align}
and we obtain
\begin{align}
\mathbb{S}(\omega) = \begin{pmatrix} \bmc{S}(\omega) & \bmc{R}(\omega) \\ \bmc{R}(\omega)^\text{H} & \bmc{A} \end{pmatrix} .
\end{align}
Since $\mathbb{S}(\omega)$ is a power spectral density matrix and thus Hermitian and positive definite, this implies that the Schur complements of its diagonal blocks are likewise positive definite,
\begin{align}
\bmc{S}(\omega) - \bmc{R}(\omega) \bmc{A}^{-1} \bmc{R}(\omega)^\text{H} \geq 0 \qquad \text{and} \qquad \bmc{A} - \bmc{R}(\omega)^\text{H} \bmc{S}(\omega)^{-1} \bmc{R}(\omega) \geq 0 . \label{fffri-2}
\end{align}
The second relation is precisely the fluctuation-response inequality (FRI), \eqref{fffri} of the main text,
\begin{align}
\bmc{R}(\omega)^\text{H} \bmc{S}(\omega)^{-1} \bmc{R}(\omega) \leq \bmc{A} . \label{app-fffri}
\end{align}
Note that the above derivation shows that this result is more general, in the sense that it also applies to Langevin equations with arbitrary, position-dependent diffusion matrices, e.~g.~for heat baths with different temperatures or spatial temperature profiles.


\subsection{Underdamped Langevin dynamics} \label{sec-fffri-under}
We consider the underdamped Langevin equations for the positions $\bm{x}(t) \in \mathbb{R}^d$ and velocities $\bm{v}(t) \in \mathbb{R}^d$,
\begin{align}
\dot{\bm{x}}(t) = \bm{v}(t), \qquad \bm{m} \bm{v}(t) = - \bm{\gamma}(\bm{x}(t)) \bm{v}(t) + \bm{f}(\bm{x}(t),\bm{v}(t)) + \epsilon \sum_{q = 1}^Q \bm{g}_q(\bm{x}(t),\bm{v}(t)) \phi_q(t) + \sqrt{2 \bm{\gamma}(\bm{x}(t)) \bm{T}(\bm{x}(t))} \bm{\xi}(t) \label{app-langevin-under} .
\end{align}
Compared to \eqref{langevin-under}, we also allow for different mass, as well as non-isotropic and non-uniform friction and temperature.
Specifically, we assume that $\bm{m}$, $\bm{\gamma}(\bm{x})$ and $\bm{T}(\bm{x})$ are diagonal matrices with positive entries; since they are diagonal, they mutually commute.
We also allow the steady-state force, as well as the perturbation force, to depend on velocity, as in the case of magnetic fields.
Formally, we can write this as an equation of the form \eqref{langevin} by defining the phase-space coordinate $\bm{u}(t) = (\bm{x}(t),\bm{v}(t))$,
\begin{align}
\dot{\bm{u}}(t) = \bm{a}_u(\bm{u}(t)) + \epsilon \sum_{q = 1}^Q \bm{y}_u(\bm{u}(t)) \phi_q(t) + \sqrt{2 \bm{B}_u(\bm{u}(t))} \cdot \bm{\xi}_u(t) .
\end{align}
However, the specific form of \eqref{app-langevin-under} leads to two crucial differences:
First, the diffusion matrix
\begin{align}
\bm{B}_u(\bm{u}) = \begin{pmatrix} \bm{0} & \bm{0} \\ \bm{0} & \bm{m}^{-2} \bm{\gamma}(\bm{x}) \bm{T}(\bm{x}) \end{pmatrix}
\end{align}
is singular.
At first glance, this means that the change in the path probability \eqref{path-log}, which involves the inverse of the diffusion matrix, is ill-defined.
However, the second difference is that the perturbation force only acts on the equations for the velocity, which are associated with the non-singular components of the diffusion matrix.
As a consequence, the expression \eqref{path-log} remains valid when replacing the inverse of the diffusion matrix with the inverse of the velocity-block of the diffusion matrix.

The remaining derivation proceeds then in the same way as in the overdamped case.
We define the observables,
\begin{align}
r_k(t) = z_k(\bm{x}(t),\bm{v}(t)) + \bm{w}_k(\bm{x}(t),\bm{v}(t)) \circ \dot{\bm{v}}(t) . \label{observable-under}
\end{align}
Since the velocities are now explicit configuration variables, it is sufficient to consider configuration-dependent observables with $\bm{w}_k = 0$ to reproduce \eqref{observable}.
The definition of the power-spectral density $\bmc{S}(\omega)$ \eqref{power-spectral-density} is completely analog to the overdamped case.
For the response of the observables, we obtain in analogy to \eqref{response-correlation}
\begin{align}
R_{kq}(t-t') = \Av{r_k(t) \psi_q(t')}_\text{st},
\end{align}
where the quantity $\psi_q(t)$ is now given by
\begin{align}
\psi_q(t) &= \frac{1}{2} \bm{g}_q(\bm{x}(t),\bm{v}(t)) \bm{\gamma}(\bm{x}(t))^{-1} \bm{T}(\bm{x}(t))^{-1}  \sqrt{2 \bm{\gamma}(\bm{x}(t)) \bm{T}(\bm{x}(t))} \cdot \bm{\xi}(t) \\
&= \frac{1}{2} \bm{g}_q(\bm{x}(t),\bm{v}(t)) \bm{\gamma}(\bm{x}(t))^{-1} \bm{T}(\bm{x}(t))^{-1} \cdot \Big( \bm{m} \dot{\bm{v}}(t) + \bm{\gamma}(\bm{x}(t)) \bm{v}(t) - \bm{f}(\bm{x}(t),\bm{v}(t))  \Big) . \n
\end{align}
Since the diffusion matrix is independent of $\bm{v}$, converting from the Ito- to the Stratonovich product now simply yields
\begin{align}
\psi_q(t) = \frac{1}{2} \bigg( \bm{g}_q(\bm{x}(t),\bm{v}(t)) & \bm{\gamma}(\bm{x}(t))^{-1} \bm{T}(\bm{x}(t))^{-1} \Big( \bm{\gamma}(\bm{x}(t)) \bm{v}(t) - \bm{f}(\bm{x}(t),\bm{v}(t)) \Big) - \bm{m}^{-1} \grad_v \bm{g}_q(\bm{x}(t),\bm{v}(t)) \bigg) \\
& + \frac{1}{2} \bm{g}_q(\bm{x}(t),\bm{v}(t)) \bm{\gamma}(\bm{x}(t))^{-1} \bm{T}(\bm{x}(t))^{-1} \circ \dot{\bm{v}}(t) \n .
\end{align}
This is again of the same type as \eqref{observable-under} and we can write the joint spectral density matrix of the observables $r_k(t)$ and the perturbation observables $\psi_q(t)$ as
\begin{align}
\mathbb{S}(\omega) = \begin{pmatrix} \bmc{S}(\omega) & \bmc{R}(\omega) \\ \bmc{R}(\omega)^\text{H} & \bmc{A} \end{pmatrix},
\end{align}
where the matrix $\bmc{A}$ is now given by
\begin{align}
\mathcal{A}_{qr} = \frac{1}{2} \Av{\bm{g}_q \bm{\gamma}^{-1} \bm{T}^{-1} \bm{g}_r}_\text{st} ,
\end{align}
Since \eqref{fffri-2} only depends on the positive-definiteness of the power spectrum, it also applies in the underdamped case and we still have \eqref{app-fffri},
\begin{align}
\bmc{R}(\omega)^\text{H} \bmc{S}(\omega)^{-1} \bmc{R}(\omega) \leq \bmc{A} \label{fffri-under} .
\end{align}
Interestingly, while the dynamics and thus the response and fluctuations depend on the masses, the matrix $\bmc{A}$ and thus the upper bound in the FRI are independent of $\bm{m}$.

\subsection{Markov jump dynamics}  \label{sec-fffri-jump}
Whereas in the previous cases, the configuration of the system was determined by a set of continuous real-valued variables, we now consider the case of a discrete set of configurations $\alpha(t) \in \lbrace 1,\ldots,N \rbrace$.
The transitions between different configurations are described by non-negative rates $\Omega(\alpha \vert \beta,t)$; the probability of observing a transition to configuration $\alpha$ during the time-interval $[t,t+dt]$, provided that the system was in configuration $\beta$ at time $t$ is $\Omega(\alpha \vert \beta,t) dt$.
The conditional probability $p(\alpha,t \vert \gamma,0)$ of observing configuration $\alpha$ at time $t$ conditioned on being in configuration $\gamma$ at time $0$ is then determined by the master equation
\begin{align}
d_t p(\alpha,t \vert \gamma,0) = \sum_{\beta = 1}^N \Big( \Omega(\alpha \vert \beta,t) p(\beta,t \vert \gamma,0) - \Omega(\beta \vert \alpha,t) p(\alpha,t \vert \gamma,0) \Big) . \label{master}
\end{align}
We write the rates as ($\alpha \neq \beta$)
\begin{align}
\Omega_\epsilon(\alpha \vert \beta,t) = \Omega_0(\alpha \vert \beta) \exp \bigg( \epsilon \sum_{q = 1}^Q Y_q(\alpha \vert \beta) \phi_q(t) \bigg) \label{rates} ,
\end{align}
with the unperturbed rates $\Omega_0(\alpha \vert \beta) \geq 0$, which ensures that the rates remain positive in the presence of the perturbation.
For $\epsilon = 0$, the rates are time-independent and we assume that the probability relaxes to a steady state $p_\text{st}(\alpha)$, determined by
\begin{align}
0 = \sum_{\beta = 1}^N \Big( \Omega_0(\alpha \vert \beta) p_\text{st}(\beta) - \Omega_0(\beta \vert \alpha) p_\text{st}(\alpha) \Big).
\end{align}
We define the observables as
\begin{align}
r_k(t) = z_k(\alpha(t)) + \frac{1}{dt} w_k(\alpha(t+dt) \vert \alpha(t)) \label{observable-jump} ,
\end{align}
where $z_k(\alpha)$ is a configuration-dependent function and the transition-dependent term $w_k(\alpha \vert \beta)$ is non-zero only if $\alpha \neq \beta$.
The steady-state average of the observable is
\begin{align}
\Av{r_k}_\text{st} = \sum_\alpha z_k(\alpha) p_\text{st}(\alpha) + \sum_{\alpha,\beta; \alpha \neq \beta} w_k(\alpha \vert \beta) \Omega_0(\alpha \vert \beta) p_\text{st}(\beta) .
\end{align}

The probability of a discretized trajectory during a time interval $[0,t]$, $\bar{\Gamma} = \lbrace \alpha(\theta dt) \rbrace_{\theta \in \lbrace 0,\ldots,M \rbrace}$, where $t = M dt$ can we written as
\begin{align}
\mathbb{P}_\epsilon(\bar{\Gamma}) = \prod_{\theta = 1}^M p_\epsilon\big(\alpha_\theta, \theta dt \vert \alpha_{\theta-1}, (\theta-1)dt \big) p_\text{st}(\alpha_0) ,
\end{align}
where we assume that the system is initially in the unperturbed steady state.
The logarithm of the path probability is then
\begin{align}
\ln \mathbb{P}_\epsilon(\bar{\Gamma}) = \sum_{\theta = 1}^M \ln p_\epsilon\big(\alpha_\theta, \theta dt \vert \alpha_{\theta-1}, (\theta-1)dt \big) + \ln p_\text{st}(\alpha_0) .
\end{align}
For a sufficiently short time-interval $dt$, the short-time transition is given by
\begin{align}
p_\epsilon\big(\alpha_\theta, \theta dt \vert \alpha_{\theta-1}, (\theta-1)dt \big) = \left\lbrace \begin{array}{ll}
1 - \sum_{\beta \neq \alpha_{\theta-1}} \Omega_\epsilon(\beta \vert \alpha_{\theta-1},(\theta - 1)dt) dt &\text{if} \; \alpha_\theta = \alpha_{\theta-1} \\[2ex]
\Omega_{\epsilon}(\alpha_\theta \vert \alpha_{\theta-1},(\theta - 1)dt) dt &\text{if} \; \alpha_\theta \neq \alpha_{\theta-1} . 
\end{array}  \right.
\end{align}
The expression in the first line follows from the second line by conservation of probability.
Taking the logarithm and the derivative with respect to $\epsilon$ at $\epsilon = 0$, we obtain
\begin{align}
\partial_\epsilon \ln p_\epsilon\big(\alpha_\theta, \theta dt \vert \alpha_{\theta-1}, (\theta-1)dt \big) \Big\vert_{\epsilon = 0} = \left\lbrace \begin{array}{ll}
- \frac{ \sum_{\beta \neq \alpha_{\theta-1}} \Omega_0(\beta \vert \alpha_{\theta-1}) \sum_{q = 1}^Q Y_q(\beta \vert \alpha_{\theta-1}) \phi_q((\theta-1)dt) dt}{1 - \sum_{\beta \neq \alpha_{\theta-1}} \Omega_0(\beta \vert \alpha_{\theta-1},(\theta - 1)dt) dt} &\text{if} \; \alpha_\theta = \alpha_{\theta-1} \\[2ex]
\sum_{q = 1}^Q Y_q(\alpha_\theta \vert \alpha_{\theta-1}) \phi_q((\theta-1)dt) &\text{if} \; \alpha_\theta \neq \alpha_{\theta-1} . 
\end{array}  \right.
\end{align}
Setting $Y_q(\alpha \vert \alpha) = 0$ and expanding for small $dt$, we can write the derivative of the trajectory probability as
\begin{gather}
\partial_\epsilon \ln \mathbb{P}_\epsilon(\bar{\Gamma}) \Big\vert_{\epsilon = 0} = \sum_{q = 1}^Q \sum_{\theta = 1}^M \bigg(\tilde{z}_q(\alpha_{\theta-1}) + \frac{1}{dt} \tilde{w}_q(\alpha_\theta \vert \alpha_{\theta-1})\bigg) \phi_q((\theta-1)dt) dt  \\
\text{with} \qquad \tilde{z}_q(\alpha) = - \sum_{\beta} \Omega_0(\beta \vert \alpha) Y_q(\beta \vert \alpha), \qquad \tilde{w}_k(\alpha \vert \beta) = Y_q(\alpha \vert \beta) \n .
\end{gather}
In the continuous-time limit, we obtain
\begin{align}
\partial_\epsilon \ln \mathbb{P}_\epsilon(\bar{\Gamma}) \Big\vert_{\epsilon = 0} = \sum_{q=1}^Q \int_0^t dt' \ \psi_q(t') \phi_q(t') \qquad
\text{with} \qquad \psi_q(t) = \tilde{z}_q(\alpha(t)) + \frac{1}{dt} \tilde{w}_q(\alpha(t+dt) \vert \alpha(t)) .
\end{align}
Just as in the Langevin case, we can thus define a set of response observables $\psi_q(t)$ of the same type as \eqref{observable-jump}, in terms of which we can write
\begin{align}
\partial_\epsilon \Av{r_k(t)}_\epsilon \Big\vert_{\epsilon = 0} = \sum_{q = 1}^Q \int_0^t dt' \ R_{kq}(t-t') \phi_q(t') \qquad \text{with} \qquad R_{kq}(t-t') = \Av{r_k(t) \psi_q(t')}_\text{st} .
\end{align}
Moreover, we have for the correlation between the response observables
\begin{align}
\Av{\psi_q(t) \psi_r(t')}_\text{st} = \Bigg\langle \bigg(\frac{1}{dt} &Y_q(\alpha(t+dt)\vert \alpha(t)) - \sum_{\beta} \Omega_0(\beta \vert \alpha(t)) Y_q(\beta \vert \alpha(t)) \bigg) \\
&\times \bigg(\frac{1}{dt} Y_r(\alpha(t'+dt)\vert \alpha(t')) - \sum_{\beta} \Omega_0(\beta \vert \alpha(t')) Y_r(\beta \vert \alpha(t')) \bigg) \Bigg\rangle_\text{st} \n .
\end{align}
Suppose that $t > t'+dt$,
In that case, since the dynamics is Markovian, the probability of being in configuration $\alpha(t)$ at time $t$ only depends on the last observed configuration $\alpha(t'+dt)$.
We can then consider the conditional average
\begin{align}
\Bigg\langle \bigg(\frac{1}{dt} &Y_q(\alpha(t+dt)\vert \alpha(t)) - \sum_{\beta} \Omega_0(\beta \vert \alpha(t)) Y_q(\beta \vert \alpha(t)) \bigg) \bigg\vert \alpha(t'+dt) = \tilde{\alpha} \Bigg\rangle_\text{st} \\
& = \sum_{\alpha_2,\alpha_1} \bigg( \frac{1}{dt} Y_q(\alpha_2 \vert \alpha_1) - \sum_{\beta} \Omega_0(\beta \vert \alpha_1) Y_q(\beta \vert \alpha_1) \bigg) p_0(\alpha_2, t+dt; \alpha_1,t \vert \tilde{\alpha}, t'+dt)  \nn
& = \sum_{\alpha_2,\alpha_1} \bigg( \frac{1}{dt} Y_q(\alpha_2 \vert \alpha_1) - \sum_{\beta} \Omega_0(\beta \vert \alpha_1) Y_q(\beta \vert \alpha_1) \bigg) p_0(\alpha_2, t+dt \vert \alpha_1,t) p_0(\alpha_1,t \vert \tilde{\alpha}, t'+dt) \n ,
\end{align}
where we once again used that the probability at time $t+dt$ only depends on the configuration at time $t$.
Since $Y_q(\alpha_2 \vert \alpha_1)$ is non-zero only for $\alpha_2 \neq \alpha_1$, we have  $p_0(\alpha_2, t+dt \vert \alpha_1,t) = \Omega_0(\alpha_2 \vert \alpha_1) dt$ and thus,
\begin{align}
\Bigg\langle \bigg(\frac{1}{dt} &Y_q(\alpha(t+dt)\vert \alpha(t)) - \sum_{\beta} \Omega_0(\beta \vert \alpha(t)) Y_q(\beta \vert \alpha(t)) \bigg) \bigg\vert \alpha(t'+dt) = \tilde{\alpha} \Bigg\rangle_\text{st} \\
& = \sum_{\alpha_1} \bigg( \sum_{\alpha_2} \Omega_0(\alpha_2 \vert \alpha_1) Y_q(\alpha_2 \vert \alpha_1)  - \sum_{\beta} \Omega_0(\beta \vert \alpha_1) Y_q(\beta \vert \alpha_1) \bigg) p_0(\alpha_1,t \vert \tilde{\alpha}, t'+dt) = 0 \n .
\end{align}
Thus, $\Av{\psi_q(t) \psi_r(t')}_\text{st} = 0$ for $t > t'$, and analogously for $t < t'$, and the only non-zero value is obtained for $t = t'$,
\begin{align}
\Av{\psi_q(t) \psi_r(t')}_\text{st} &= \Bigg\langle \bigg(\frac{1}{dt} Y_q(\alpha(t+dt)\vert \alpha(t)) - \sum_{\beta} \Omega_0(\beta \vert \alpha(t)) Y_q(\beta \vert \alpha(t)) \bigg) \\
&\hspace{2cm} \times \bigg(\frac{1}{dt} Y_r(\alpha(t+dt)\vert \alpha(t)) - \sum_{\beta} \Omega_0(\beta \vert \alpha(t)) Y_r(\beta \vert \alpha(t)) \bigg) \Bigg\rangle_\text{st} \nn
&= \sum_{\alpha_2} \sum_{\alpha_1} \bigg( \frac{1}{dt} Y_q(\alpha_2 \vert \alpha_1) - \sum_{\beta} \Omega_0(\beta \vert \alpha_1) Y_q(\beta \vert \alpha_1) \bigg) \bigg( \frac{1}{dt} Y_r(\alpha_2 \vert \alpha_1) - \sum_{\beta} \Omega_0(\beta \vert \alpha_1) Y_r(\beta \vert \alpha_1) \bigg)  \nn
&\hspace{3cm} \times p_0(\alpha_2, t+dt \vert \alpha_1,t) p_\text{st}(\alpha_1) \nn
&\simeq \frac{1}{dt} \sum_{\alpha_2} \sum_{\alpha_1} Y_q(\alpha_2 \vert \alpha_1) Y_r(\alpha_2 \vert \alpha_1) \Omega_0(\alpha_2 \vert \alpha_1) p_\text{st}(\alpha_1) + O(dt^0) , \n
\end{align}
where in the last line we only keep the leading-order term in $dt$.
In the continuous-time limit, we thus have
\begin{align}
\Av{\psi_q(t) \psi_r(t')}_\text{st} = \sum_{\alpha_2} \sum_{\alpha_1} Y_q(\alpha_2 \vert \alpha_1) Y_r(\alpha_2 \vert \alpha_1) \Omega_0(\alpha_2 \vert \alpha_1) p_\text{st}(\alpha_1) \delta(t-t') ,
\end{align}
in analogy to \eqref{response-observable-correlation}.
The delta-correlations of the response observable are a direct consequence of the Markovianity of the dynamics---white Gaussian noise in the Langevin case and white Poisson noise in the jump case.

The remainder of the derivation proceeds analog to the Langevin case and we obtain
\begin{align}
\bmc{R}(\omega)^\text{H} \bmc{S}(\omega)^{-1} \bmc{R}(\omega) \leq \bmc{A} \qquad \text{with} \qquad \mathcal{A}_{qr} = \sum_{\alpha,\beta} Y_q(\alpha \vert \beta) Y_r(\alpha \vert \beta) \Omega_0(\alpha \vert \beta) p_\text{st}(\beta) . \label{fffri-jump}
\end{align}
Note that if the perturbations are either even or odd under time-reversal, $Y_q(\alpha \vert \beta) = \pm Y_q(\beta \vert \alpha)$, we can write the matrix $\bmc{A}$ as
\begin{align}
\mathcal{A}_{qr} = \sum_{\alpha,\beta} Y_q(\alpha \vert \beta) Y_r(\alpha \vert \beta) \mathcal{T}(\alpha, \beta)  \qquad \text{with} \qquad \mathcal{T}_\text{st}(\alpha, \beta) = \frac{1}{2} \big( \Omega_0(\alpha \vert \beta) p_\text{st}(\beta) + \Omega_0(\beta \vert \alpha) p_\text{st}(\alpha) \big) .
\end{align}
The quantity $\mathcal{T}_\text{st}(\alpha, \beta)$ is called the (steady-state) traffic between $\alpha$ and $\beta$; it measures the average rate of transitions between the configurations in either direction.
The sum of the traffic over all pairs of configurations is called the activity; thus, $\bmc{A}$ measures the activity in the unperturbed system weighted by the perturbations.

A particular case of \eqref{rates} which assigns an energetic interpretation to the rates is
\begin{align}
\Omega_\epsilon(\alpha \vert \beta) = \Omega_0 \exp \bigg( - \frac{\beta}{2} \Big( B(\alpha,\beta) + U(\alpha) - U(\beta) - A(\alpha,\beta) + \epsilon G(\alpha,\beta) \phi(t) \Big) \bigg) .
\end{align}
Here, $\Omega_0$ is an overall rate and $\beta = 1/T$ is the inverse temperature.
The symmetric energy $B(\alpha,\beta) = B(\beta,\alpha)$ corresponds to an energy barrier between configurations $\alpha$ and $\beta$.
$U(\alpha)$ is the energy assigned to configuration $\alpha$ and $A(\alpha,\beta) = -A(\beta,\alpha)$ measures the energy supplied to the system by an external, non-conservative force in the transition from $\beta$ to $\alpha$.
In this case, the perturbation $G(\alpha,\beta)$ corresponds to changing either the heights of the barriers (if $G(\alpha,\beta) = G(\beta,\alpha)$), the energies of the configurations (if $G(\alpha,\beta) = V(\alpha) - V(\beta)$) or the energy supplied by the non-conservative force (if $G(\alpha,\beta)$ is antisymmetric but cannot be written as an energy difference).

\section{Linear Langevin dynamics} \label{sec-linear}

\subsection{Overdamped} \label{sec-linear-over}
Let us consider the special case of \eqref{langevin}
\begin{align}
\dot{\bm{x}}(t) = -\bm{A} \big(\bm{x}(t) - \bm{a} \big) + \epsilon \sum_{q=1}^Q \bm{y}_q \phi_q(t) + \sqrt{2 \bm{B}} \bm{\xi}(t). \label{langevin-linear-over}
\end{align}
Here, the force is linear in the coordinates, with a matrix $\bm{A}$ that has eigenvalues with positive real part to ensure that the steady state is stable.
The diffusion matrix $\bm{B}$ is constant and positive definite and the perturbations are constant vectors $\bm{y}_q$.
Since this equation is linear, the steady state probability density is Gaussian,
\begin{align}
p_\text{st}(\bm{x}) = \frac{1}{ \sqrt{(2 \pi)^D \det(\bm{\Xi}_\text{st})}} \exp \bigg( - \frac{1}{2} \big( \bm{x} - \bm{a} \big) \bm{\Xi}_\text{st}^{-1} \big( \bm{x} - \bm{a} \big) \bigg) ,
\end{align}
where the steady-state covariance matrix $(\bm{\Xi}_\text{st})_{jh} = \text{Cov}_\text{st}(x_j,x_h)$ is determined from the Lyapunov equation
\begin{align}
\bm{A} \bm{\Xi}_\text{st} + \bm{\Xi}_\text{st} \bm{A}^\text{T} = 2 \bm{B} . \label{lyapunov}
\end{align}
The system is in equilibrium if $\bm{A} \bm{B} = \bm{B} \bm{A}^\text{T}$, in which case $\bm{\Xi}_\text{st} = \bm{A}^{-1} \bm{B}$.
For given $\bm{x}(t')$, we can calculate the conditional average for $t > t'$ in the unperturbed system,
\begin{align}
\Av{\bm{x}(t) \vert \bm{x}(t')} = \bm{a} + e^{-\bm{A}(t-t')} \big( \bm{x}(t') - \bm{a} \big) .
\end{align}
Multiplying by $\bm{x}(t') - \bm{a}$ and averaging with respect to the steady-state density, we find
\begin{align}
\bm{C}(t-t') = \text{Cov}_\text{st}\big(\bm{x}(t),\bm{x}(t') \big) = \left\lbrace \begin{array}{ll}
e^{-\bm{A}(t-t')} \bm{\Xi}_\text{st} &\text{for} \; t \geq t' \\[1 ex]
\bm{\Xi}_\text{st} e^{-\bm{A}^\text{T}(t'-t)} & \text{for} \; t' > t .
\end{array}  \right.
\end{align}
Taking the Fourier transform, we obtain the power-spectral density matrix
\begin{align}
\bmc{S}(\omega) &= \big( \bm{A} - i \omega \bm{I}\big)^{-1} \bm{\Xi}_\text{st} + \bm{\Xi}_\text{st} \big( \bm{A}^\text{T} + i \omega \bm{I}\big)^{-1} \label{spectrum-linear-over} \\
&= \big( \bm{A}^2 + \omega^2 \bm{I}\big)^{-1} \bm{A} \bm{\Xi}_\text{st} + \bm{\Xi}_\text{st} \bm{A}^\text{T} \big( \bm{A}^{2,\text{T}} + \omega^2 \bm{I}\big)^{-1} + i \omega \Big( \big( \bm{A}^2 + \omega^2 \bm{I}\big)^{-1} \bm{\Xi}_\text{st} - \bm{\Xi}_\text{st} \big( \bm{A}^{2,\text{T}} + \omega^2 \bm{I}\big)^{-1} \Big) . \n
\end{align}
In equilibrium, the imaginary part vanishes, while the real part is the symmetric matrix
\begin{align}
\bmc{S}(\omega) &= \big( \bm{A}^2 + \omega^2 \bm{I}\big)^{-1} \bm{B} + \bm{B} \big( \bm{A}^{2} + \omega^2 \bm{I}\big)^{-1,\text{T}} .
\end{align}
In the presence of the perturbation, the average of $\bm{x}(t)$ is given by
\begin{align}
\Av{\bm{x}(t)}_\epsilon - \Av{\bm{x}}_\text{st} = \epsilon \sum_{q=1}^Q \int_0^t dt' \ e^{-\bm{A}(t-t')} \bm{y}_q \phi_q(t'),
\end{align}
and we can identify the response matrix of the positions for $t \geq 0$
\begin{align}
R_{jq}(t) = \big(e^{-\bm{A} t} \bm{y}_q \big)_j .
\end{align}
We choose $\bm{y}_q = y_0 \hat{\bm{e}}_q$ where $\hat{\bm{e}}_q$ is the unit vector in the direction of $x_q$ for $q = 1,\ldots,D$, which yields
\begin{align}
\bm{R}(t) = y_0 e^{-\bm{A} t} \qquad \Leftrightarrow \qquad \bmc{R}(\omega) = y_0 \big(\bm{A} - i \omega \bm{I} \big)^{-1} \label{response-linear-over} .
\end{align}
Combining \eqref{spectrum-linear-over} and \eqref{response-linear-over}, we find
\begin{align}
\bmc{R}(\omega)^\text{H} \bmc{S}(\omega)^{-1} \bmc{R}(\omega) &= y_0^2 \big(\bm{A}^\text{T} + i \omega \bm{I} \big)^{-1} \Big( \big( \bm{A} - i \omega \bm{I}\big)^{-1} \bm{\Xi}_\text{st} + \bm{\Xi}_\text{st} \big( \bm{A}^\text{T} + i \omega \bm{I}\big)^{-1} \Big)^{-1} \big(\bm{A} - i \omega \bm{I} \big)^{-1} \\
&= y_0^2 \Big( \bm{\Xi}_\text{st} \big( \bm{A}^\text{T} + i \omega \bm{I}\big) + \big( \bm{A} - i \omega \bm{I}\big) \bm{\Xi}_\text{st} \Big)^{-1} \nn
&= \frac{1}{2} y_0^2 \bm{B}^{-1} \n ,
\end{align}
where we used \eqref{lyapunov}.
This is precisely the matrix $\bmc{A}$ on the right-hand side of \eqref{app-fffri} for the perturbations $\bm{y}_q = y_0 \hat{\bm{e}}_q$,
\begin{align}
\mathcal{A}_{qr} = \frac{1}{2} y_0^2 \hat{\bm{e}}_q \bm{B}^{-1} \hat{\bm{e}}_r = \frac{1}{2} y_0^2 \big(\bm{B}^{-1}\big)_{qr} .
\end{align}
Thus, for linear dynamics, the inequality \eqref{app-fffri} turns into an equality, provided that we choose the degrees of freedom of the Langevin equation as the observables and that the perturbations are constant forces in the respective direction.
We also obtain
\begin{align}
\bmc{R}(\omega) \bmc{A}^{-1} \bmc{R}(\omega)^\text{H} &= 2 \big(\bm{A} - i \omega \bm{I} \big)^{-1} \bm{B} \big(\bm{A}^\text{T} + i \omega \bm{I} \big)^{-1} \label{response-equality-3} \\
&= \big(\bm{A} - i \omega \bm{I} \big)^{-1} \big( \bm{A} \bm{\Xi}_\text{st} + \bm{\Xi}_\text{st} \bm{A}^\text{T} \big) \big(\bm{A}^\text{T} + i \omega \bm{I} \big)^{-1} \nn
&= \big(\bm{A} - i \omega \bm{I} \big)^{-1} \Big( \big(\bm{A} - i \omega \bm{I} \big) \bm{\Xi}_\text{st} + \bm{\Xi}_\text{st} \big(\bm{A}^\text{T} + i \omega \bm{I} \big) \Big) \big(\bm{A}^\text{T} + i \omega \bm{I} \big)^{-1} \nn
&= \big( \bm{A} - i \omega \bm{I}\big)^{-1} \bm{\Xi}_\text{st} + \bm{\Xi}_\text{st} \big( \bm{A}^\text{T} + i \omega \bm{I}\big)^{-1} = \bmc{S}(\omega) \n ,
\end{align}
so that the first inequality in \eqref{fffri-2} is likewise an equality.
More generally, we consider observables that are linear combinations of the positions, $\bm{z}(t) = \bmc{Z} \bm{x}(t)$, where $\bmc{Z} \in \mathbb{R}^{K \times D}$.
For general perturbations $\bm{y}_q$, we also define the matrix $\bmc{Y} \in \mathbb{R}^{D \times Q}$ that has $\bm{y}_q$ as its $q$-th column.
Then, the corresponding power-spectral and response matrices are given by
\begin{align}
\bmc{S}(\omega) = \bmc{Z} \big( \bm{A} - i \omega \bm{I}\big)^{-1} \bm{\Xi}_\text{st} + \bm{\Xi}_\text{st} \big( \bm{A}^\text{T} + i \omega \bm{I}\big)^{-1} \bmc{Z}^\text{T}, \qquad \bmc{R}(\omega) = \bmc{Z} \big(\bm{A} - i \omega \bm{I} \big)^{-1} \bmc{Y}, \qquad \bmc{A} = \frac{1}{2} \bmc{Y}^\text{T} \bm{B}^{-1} \bmc{Y} .
\end{align}
Comparing this to the above, we find that, if $K = D$ and $\bmc{Z}$ is invertible, then
\begin{align}
\bmc{R}(\omega)^\text{H} \bmc{S}(\omega)^{-1} \bmc{R}(\omega) = \bmc{A},
\end{align}
while if $Q = D$ and $\bmc{Y}$ is invertible, then
\begin{align}
\bmc{R}(\omega) \bmc{A}^{-1} \bmc{R}(\omega)^\text{H} = \bmc{S}(\omega) .
\end{align}
Intuitively, if $\bmc{Z}$ is invertible, this means that the observables contain all information about the positions of the system, and measuring their responses and joint fluctuations can recover the magnitude of arbitrary perturbations that is characterized by the matrix $\bmc{A}$.
On the other hand, if $\bmc{Y}$ is invertible, the perturbations can reproduce any constant-in-space force and we can use them to reconstruct the fluctuations of arbitrary observables.
We stress that these results are valid independent of whether the system is in equilibrium or not; rather, they only rely on the linearity of the dynamics.
\eqref{linear-identity} of the main text is recovered for $\bm{A} = \frac{1}{\gamma} \bm{K}$, $\bm{B} = \frac{T}{\gamma} \bm{I}$ and $\bmc{Y} = \frac{1}{\gamma} \bmc{G}$, where $\bm{K}$ has the dimensions of a spring constant and $\bmc{G}$ of a force.

\subsection{Underdamped} \label{sec-linear-under}
Next, we consider the linear underdamped equations
\begin{align}
\dot{\bm{x}}(t) = \bm{v}(t), \qquad \bm{m} \dot{\bm{v}}(t) = - \bm{\gamma} \bm{v}(t) - \bm{K} \big(\bm{x}(t) - \bm{a} \big) + \epsilon \sum_{q=1}^Q \bm{y}_q \phi_q(t) + \sqrt{2 \bm{\gamma} \bm{T}} \bm{\xi}(t),
\end{align}
where $\bm{m}$, $\bm{\gamma}$ and $\bm{T}$ are diagonal matrices with positive entries and $\bm{K}$ is a matrix whose eigenvalues have positive real parts.
As in Section \ref{sec-fffri-under}, we can formally write this as a linear overdamped equation with singular diffusion matrix by introducing the phase-space variable $\bm{u}(t) = (\bm{x}(t),\bm{v}(t))$,
\begin{gather}
\dot{\bm{u}}(t) = - \bm{A}_u \big( \bm{u}(t) - \bm{a}_u \big) + \epsilon \sum_{q=1}^Q \bm{y}_{u,q} \phi_q(t) + \sqrt{2 \bm{B}_u} \bm{\xi}(t) \qquad \text{with} \\
\bm{A}_u = \begin{pmatrix} \bm{0} & - \bm{I} \\ \bm{m}^{-1} \bm{K} & \bm{m}^{-1} \bm{\gamma} \end{pmatrix}, \qquad \bm{B}_u = \begin{pmatrix} \bm{0} & \bm{0} \\ \bm{0} & \bm{m}^{-2} \bm{\gamma} \bm{T} \end{pmatrix}, \qquad \bm{a}_u = \begin{pmatrix} \bm{a} \\ \bm{0} \end{pmatrix}, \qquad \bm{y}_{u,q} = \begin{pmatrix} \bm{0} \\ \bm{m}^{-1} \bm{y}_q \end{pmatrix} \n .
\end{gather}
In analogy to \eqref{spectrum-linear-over}, the power-spectral density matrix of the phase-space variables is given by
\begin{align}
\bmc{S}_u(\omega) = \big( \bm{A}_u - i \omega \bm{I}\big)^{-1} \bm{\Xi}_{u,\text{st}} + \bm{\Xi}_{u,\text{st}} \big( \bm{A}_u^\text{T} + i \omega \bm{I}\big)^{-1} \label{spectral-linear-under}.
\end{align}
From this, we can obtain the power-spectral density of the position degrees of freedom as the upper left block,
\begin{align}
\bmc{S}_x(\omega) = \big(\bmc{S}_u(\omega) \big)_{xx}
\end{align}
Since the perturbations only act on the equation for $\bm{v}$, we further have
\begin{align}
\bmc{R}_u(\omega) = \begin{pmatrix} \big(( \bm{A}_u - i \omega \bm{I})^{-1} \big)_{xv} \bm{m}^{-1} \bmc{Y} \\ \big(( \bm{A}_u - i \omega \bm{I})^{-1} \big)_{vv} \bm{m}^{-1} \bmc{Y}  \end{pmatrix}, \label{response-linear-under}
\end{align}
where, as in the overdamped case the matrix $\bmc{Y}$ has $\bm{y}_q$ as its $q$-th column.
The first component is the response $\bmc{R}_x(\omega)$ of the position degrees of freedom.
To proceed, we note that, due to the deterministic relation $\dot{\bm{x}}(t) = \bm{v}(t)$, which, in the Fourier domain becomes $-i \omega \hat{\bm{x}}(\omega) = \hat{\bm{v}}(\omega)$, the velocity and position components of $\bmc{S}_u(\omega)$ and $\bmc{R}_u(\omega)$ are related to each other by
\begin{align}
\big(\bmc{S}_u(\omega) \big)_{vv} = \omega^2 \big(\bmc{S}_u(\omega) \big)_{xx}, \qquad \big(\bmc{S}_u(\omega) \big)_{vx} = -i \omega \big(\bmc{S}_u(\omega) \big)_{xx},   \qquad \big( \bmc{R}_u(\omega) \big)_v = - i \omega \big( \bmc{R}_u(\omega) \big)_x .
\end{align}
Thus, it is sufficient to determine the fluctuations and response of either the position or the velocity; in the following, we will focus on the velocity.
We can evaluate this further by expressing the inverse of the matrix $\bm{A}_u - i \omega \bm{I}$ in terms of its blocks,
\begin{align}
\big(\bm{A}_u - i \omega \bm{I} \big)^{-1} = \begin{pmatrix} 
\big( ( \bm{\gamma} - i \omega \bm{m} )^{-1} \bm{K} - i \omega \bm{I} \big)^{-1} 
& \big( \bm{m}^{-1} \bm{K} - i \omega \bm{m}^{-1} (\bm{\gamma} - i \omega \bm{m}) \big)^{-1} \\ 
- \frac{i}{\omega} \big( -(i \omega \bm{m})^{-1} \bm{K} + \bm{m}^{-1}(\bm{\gamma} - i \omega \bm{m}) \big)^{-1} \bm{m}^{-1} \bm{K}  
& \big( -(i \omega \bm{m})^{-1} \bm{K} + \bm{m}^{-1}(\bm{\gamma} - i \omega \bm{m}) \big)^{-1}  \end{pmatrix},
\end{align}
which results in
\begin{align}
\bmc{R}_v(\omega) = - i \omega \big( \bm{m}^{-1} \bm{K} - i \omega \bm{m}^{-1} (\bm{\gamma} - i \omega \bm{m}) \big)^{-1} \bm{m}^{-1} \bmc{Y} .
\end{align}
The matrix $\bmc{A}$ follows directly from \eqref{fffri-under},
\begin{align}
\bmc{A} = \frac{1}{2} \bmc{Y}^\text{T} \big(\bm{\gamma} \bm{T}\big)^{-1} \bmc{Y} .
\end{align}
If $\bmc{Y}$ is invertible, then
\begin{align}
\bmc{R}_v(\omega) \bmc{A}^{-1} \bmc{R}_v(\omega)^\text{H} = 2 \omega^2 \big( \bm{m}^{-1} \bm{K} - i \omega \bm{m}^{-1} (\bm{\gamma} - i \omega \bm{m}) \big)^{-1} \bm{m}^{-1} \bm{\gamma} \bm{T} \bm{m}^{-1} \big(  \bm{K}^\text{T} \bm{m}^{-1} + i \omega \bm{m}^{-1} (\bm{\gamma} + i \omega \bm{m}) \big)^{-1} \label{response-equality-1} .
\end{align}
We want to show that this is the same as the velocity component of the spectral density matrix,
\begin{align}
\bmc{S}_v(\omega) &= \big(\bm{m}^{-1} \bm{K} - i \omega \bm{m}^{-1}(\bm{\gamma} - i \omega \bm{m}) \big)^{-1} \big( - i \omega \bm{\Xi}_{vv,\text{st}} - \bm{m}^{-1} \bm{K} \bm{\Xi}_{xv,\text{st}} \big) \label{response-equality-2} \\
&\qquad + \big(i \omega \bm{\Xi}_{vv,\text{st}} - \bm{\Xi}_{vx,\text{st}} \bm{K}^\text{T} \bm{m}^{-1} \big) \big( \bm{K}^\text{T} \bm{m}^{-1}  + i \omega \bm{m}^{-1}(\bm{\gamma} + i \omega \bm{m}) \big)^{-1} \nn
&= \big(\bm{m}^{-1} \bm{K} - i \omega \bm{m}^{-1}(\bm{\gamma} - i \omega \bm{m}) \big)^{-1} \nn
&\qquad \times \bigg[ \big( - i \omega \bm{\Xi}_{vv,\text{st}} - \bm{m}^{-1} \bm{K} \bm{\Xi}_{xv,\text{st}} \big) \big( \bm{K}^\text{T} \bm{m}^{-1}  + i \omega \bm{m}^{-1}(\bm{\gamma} + i \omega \bm{m}) \big) \nn
&\qquad \qquad + \big(\bm{m}^{-1} \bm{K} - i \omega \bm{m}^{-1}(\bm{\gamma} - i \omega \bm{m}) \big) \big(i \omega \bm{\Xi}_{vv,\text{st}} - \bm{\Xi}_{vx,\text{st}} \bm{K}^\text{T} \bm{m}^{-1} \big) \bigg] \nn
&\qquad \times \big( \bm{K}^\text{T} \bm{m}^{-1}  + i \omega \bm{m}^{-1}(\bm{\gamma} + i \omega \bm{m}) \big)^{-1} . \n
\end{align}
Evaluating the term in square brackets, we find
\begin{align}
\bigg[ \ldots \bigg] &= -i \omega \Big( \bm{m}^{-1} \bm{K} \big( -\bm{\Xi}_{vv,\text{st}} + \bm{\Xi}_{xv} \bm{m}^{-1} \bm{\gamma} \big) - \big( -\bm{\Xi}_{vv,\text{st}} + \bm{\gamma}  \bm{m}^{-1} \bm{\Xi}_{xv} \big) \bm{K}^\text{T} \bm{m}^{-1} \Big) \label{brackets-term} \\
& \qquad + \omega^2 \Big( \bm{m}^{-1} \bm{\gamma} \bm{\Xi}_{vv,\text{st}} + \bm{\Xi}_{vv,\text{st}} \bm{\gamma} \bm{m}^{-1} + \bm{m}^{-1} \bm{K} \bm{\Xi}_{xv} + \bm{\Xi}_{vx,\text{st}} \bm{m}^{-1} \Big) - \bm{m}^{-1} \bm{K} \big( \bm{\Xi}_{xv,\text{st}} + \bm{\Xi}_{vx,\text{st}} \big) \bm{K}^\text{T} \bm{m}^{-1} \n .
\end{align}
We note that the phase-space covariance matrix $\bm{\Xi}_{u,\text{st}}$ is determined from the Lyapunov equation
\begin{align}
\bm{A}_u \bm{\Xi}_{u,\text{st}} + \bm{\Xi}_{u,\text{st}} \bm{A}_u^\text{T} = 2 \bm{B}_u ,
\end{align}
which we can write in block-wise form as
\begin{align}
&\begin{pmatrix}
- \bm{\Xi}_{vx,\text{st}} - \bm{\Xi}_{xv,\text{st}} & - \bm{\Xi}_{vv,\text{st}}  + \bm{\Xi}_{xv,\text{st}} \bm{m}^{-1} \bm{\gamma} + \bm{\Xi}_{xx,\text{st}} \bm{K}^\text{T} \bm{m}^{-1} \\
- \bm{\Xi}_{vv,\text{st}} + \bm{m}^{-1} \bm{\gamma} \bm{\Xi}_{vx,\text{st}} + \bm{m}^{-1} \bm{K} \bm{\Xi}_{xx,\text{st}}  & \bm{m}^{-1} \bm{K} \bm{\Xi}_{xv,\text{st}} + \bm{\Xi}_{vx,\text{st}} \bm{K}^\text{T} \bm{m}^{-1} + \bm{m}^{-1} \bm{\gamma} \bm{\Xi}_{vv,\text{st}} + \bm{\Xi}_{vv,\text{st}} \bm{\gamma} \bm{m}^{-1}
\end{pmatrix}  \nn
&\hspace{2cm} = \begin{pmatrix}
\bm{0} & \bm{0} \\ \bm{0} & 2 \bm{m}^{-2} \bm{\gamma} \bm{T} 
\end{pmatrix} . \label{lyapunov-under}
\end{align}
From the $xx$-component, we find $\bm{\Xi}_{vx,\text{st}} = - \bm{\Xi}_{xv,\text{st}}$.
Inserting the $xv$ and $vx$ components into \eqref{brackets-term}, we obtain
\begin{align}
\bigg[ \ldots \bigg] &= i \omega \Big( \bm{m}^{-1} \bm{K} \bm{\Xi}_{xx,\text{st}} \bm{K}^\text{T} \bm{m}^{-1} - \bm{m}^{-1} \bm{K} \bm{\Xi}_{xx,\text{st}} \bm{K}^\text{T} \bm{m}^{-1} \Big) \label{brackets-term-2} \\
& \qquad + \omega^2 \Big( \bm{m}^{-1} \bm{\gamma} \bm{\Xi}_{vv,\text{st}} + \bm{\Xi}_{vv,\text{st}} \bm{\gamma} \bm{m}^{-1} + \bm{m}^{-1} \bm{K} \bm{\Xi}_{xv} + \bm{\Xi}_{vx,\text{st}} \bm{m}^{-1} \Big) \nn
&= \omega^2 \Big( \bm{m}^{-1} \bm{\gamma} \bm{\Xi}_{vv,\text{st}} + \bm{\Xi}_{vv,\text{st}} \bm{\gamma} \bm{m}^{-1} + \bm{m}^{-1} \bm{K} \bm{\Xi}_{xv} + \bm{\Xi}_{vx,\text{st}} \bm{m}^{-1} \Big) \n  .
\end{align}
Inserting the $vv$-component into \eqref{response-equality-1}, we obtain
\begin{align}
\bmc{R}_v(\omega) &\bmc{A}^{-1} \bmc{R}_v(\omega)^\text{H} = \omega^2 \big( \bm{m}^{-1} \bm{K} - i \omega \bm{m}^{-1} (\bm{\gamma} - i \omega \bm{m})^{-1} \big)^{-1} \\
&\times \big( \bm{m}^{-1} \bm{K} \bm{\Xi}_{xv,\text{st}} + \bm{\Xi}_{vx,\text{st}} \bm{K}^\text{T} \bm{m}^{-1} + \bm{m}^{-1} \bm{\gamma} \bm{\Xi}_{vv,\text{st}} + \bm{\Xi}_{vv,\text{st}} \bm{\gamma} \bm{m}^{-1} \big) \big(  \bm{K}^\text{T} \bm{m}^{-1} + i \omega \bm{m}^{-1} (\bm{\gamma} + i \omega \bm{m})^{-1} \big)^{-1} . \n
\end{align}
Comparing this to \eqref{response-equality-2}, we indeed find
\begin{align}
\bmc{R}_v(\omega) \bmc{A}^{-1} \bmc{R}_v(\omega)^\text{H} = \bmc{S}_v(\omega) ,
\end{align}
and thus equality in the first equation of \eqref{fffri-2}.
Due to the one-to-one relation between $\bm{x}(t)$ and $\bm{v}(t)$, this translates into the same relation for the position degrees of freedom,
\begin{align}
\bmc{R}_x(\omega) \bmc{A}^{-1} \bmc{R}_x(\omega)^\text{H} = \bmc{S}_x(\omega) .
\end{align}
Finally, let us compute 
\begin{align}
\bmc{R}_v&(\omega)^\text{H} \bmc{S}_v(\omega)^{-1} \bmc{R}_v(\omega) = \omega^2 \bmc{Y}^\text{T} \bm{m}^{-1} \big(  \bm{K}^\text{T} \bm{m}^{-1} + i \omega \bm{m}^{-1} (\bm{\gamma} + i \omega \bm{m}) \big)^{-1} \\
&\times \bigg( \omega^2 \big(\bm{m}^{-1} \bm{K} - i \omega \bm{m}^{-1}(\bm{\gamma} - i \omega \bm{m}) \big)^{-1}  \Big( \bm{m}^{-1} \bm{\gamma} \bm{\Xi}_{vv,\text{st}} + \bm{\Xi}_{vv,\text{st}} \bm{\gamma} \bm{m}^{-1} + \bm{m}^{-1} \bm{K} \bm{\Xi}_{xv} + \bm{\Xi}_{vx,\text{st}} \bm{m}^{-1} \Big) \nn
&\qquad \times \big( \bm{K}^\text{T} \bm{m}^{-1}  + i \omega \bm{m}^{-1}(\bm{\gamma} + i \omega \bm{m}) \big)^{-1} \bigg)^{-1} \big( \bm{m}^{-1} \bm{K} - i \omega \bm{m}^{-1} (\bm{\gamma} - i \omega \bm{m}) \big)^{-1} \bm{m}^{-1} \bmc{Y} \n,
\end{align}
where we used \eqref{response-equality-2} and \eqref{brackets-term-2} to express the spectral density matrix.
Again using the $vv$-component of \eqref{lyapunov-under}, we obtain
\begin{align}
\bmc{R}_v(\omega)^\text{H} \bmc{S}_v(\omega)^{-1} \bmc{R}_v(\omega) = \bmc{Y}^\text{T} \bm{m}^{-1} \big( 2 \bm{m}^{-2} \bm{\gamma} \bm{T} \big)^{-1} \bm{m}^{-1} \bmc{Y} = \frac{1}{2} \bmc{Y}^\text{T} \big( \bm{\gamma} \bm{T} \big)^{-1} \bmc{Y} = \bmc{A},
\end{align}
and thus equality in the second equation of \eqref{fffri-2}.
This implies the same relation for the position degrees of freedom
\begin{align}
\bmc{R}_x(\omega)^\text{H} \bmc{S}_x(\omega)^{-1} \bmc{R}_x(\omega) = \bmc{A} .
\end{align}
For more general linear observables $\bm{z} = \bmc{Z} \bm{x}$ or $\bm{z} = \bmc{Z} \bm{v}$, the same relation holds if $\bmc{Z}$ is invertible.

\section{Local perturbations}

Let us return to the overdamped equation \eqref{langevin-linear-over} and divide the degrees of freedom into two disjoint subsets $\bm{x}(t) = (\bm{x}_a(t),\bm{x}_p(t))$.
We assume that the diffusion matrix is block-diagonal
\begin{align}
\bm{B} = \begin{pmatrix} \bm{B}_a & \bm{0} \\ \bm{0} & \bm{B}_p \end{pmatrix}
\end{align}
We consider the case where the perturbation only acts on the \enquote{active} degrees of freedom $\bm{x}_a$, that is, $\bm{y}_q$ has non-zero entries only on the $\bm{x}_a$ directions.
We also measure the response and fluctuations of the active degrees of freedom.
Writing the coupling matrix in block form
\begin{align}
\bm{A} = \begin{pmatrix} \bm{A}_{aa} & \bm{A}_{ap} \\ \bm{A}_{pa} & \bm{A}_{pp} \end{pmatrix},
\end{align}
we can write the response matrix as
\begin{align}
\bmc{R}_{a}(\omega) = \Big( \bm{A}_{aa} - i \omega \bm{I} - \bm{A}_{ap} \big( \bm{A}_{pp} - i \omega \bm{I} \big)^{-1} \bm{A}_{pa} \Big)^{-1} \bmc{Y} \equiv \bmc{Q}_{aa}(\omega) \bmc{Y}  \label{response-block-inverse}.
\end{align}
On the other hand, the $aa$ component of the power-spectral density matrix can be written as (see \eqref{response-equality-3}) 
\begin{align}
\bmc{S}_{aa}(\omega) &= 2 \big(( \bm{A} - i \omega \bm{I} )^{-1} \big)_{aa} \bm{B}_a \big(( \bm{A}^\text{T} + i \omega \bm{I} )^{-1} \big)_{aa} + 2 \big(( \bm{A} - i \omega \bm{I} )^{-1} \big)_{ap} \bm{B}_p \big(( \bm{A}^\text{T} + i \omega \bm{I} )^{-1} \big)_{pa} \\
&= 2 \bmc{Q}_{aa}(\omega) \Big( \bm{B}_a + \bm{A}_{ap} \big(\bm{A}_{pp} - i \omega \bm{I} \big)^{-1} \bm{B}_p \big(\bm{A}_{pp}^\text{T} + i \omega \bm{I} \big)^{-1} \bm{A}_{ap}^\text{T} \Big) \bmc{Q}_{aa}(\omega)^\text{H} \n .
\end{align}
We then obtain
\begin{align}
\bmc{R}_a(\omega)^\text{H} \bmc{S}_{aa}(\omega)^{-1} \bmc{R}_a(\omega) = \frac{1}{2} \bmc{Y}^\text{T} \Big( \bm{B}_a + \bm{A}_{ap} \big(\bm{A}_{pp} - i \omega \bm{I} \big)^{-1} \bm{B}_p \big(\bm{A}_{pp}^\text{T} + i \omega \bm{I} \big)^{-1} \bm{A}_{ap}^\text{T} \Big)^{-1} \bmc{Y} \label{response-equality-4} .
\end{align}
The inequality \eqref{app-fffri} follows by noting that the second term in the expression in parentheses is positive definite and thus
\begin{align}
\frac{1}{2} \bmc{Y}^\text{T} \Big( \bm{B}_a + \bm{A}_{ap} \big(\bm{A}_{pp} - i \omega \bm{I} \big)^{-1} \bm{B}_p \big(\bm{A}_{pp}^\text{T} + i \omega \bm{I} \big)^{-1} \bm{A}_{ap}^\text{T} \Big)^{-1} \bmc{Y} \leq \frac{1}{2} \bmc{Y}^\text{T} \bm{B}_a^{-1} \bmc{Y} = \bmc{A} .
\end{align}
More explicitly, we have from the Woodbury matrix identity
\begin{align}
\Big( &\bm{B}_a + \bm{A}_{ap} \big(\bm{A}_{pp} - i \omega \bm{I} \big)^{-1} \bm{B}_p \big(\bm{A}_{pp}^\text{T} + i \omega \bm{I} \big)^{-1} \bm{A}_{ap}^\text{T} \Big)^{-1} \label{woodbury} \\ 
&= \bm{B}_a^{-1} - \bm{B}_a^{-1} \bm{A}_{ap} \big(\bm{A}_{pp} - i \omega \bm{I} \big)^{-1} \Big( \bm{B}_p^{-1} + \big(\bm{A}_{pp}^\text{T} + i \omega \bm{I} \big)^{-1} \bm{A}_{ap}^\text{T} \bm{B}_a^{-1} \bm{A}_{ap}\big(\bm{A}_{pp} - i \omega \bm{I} \big)^{-1} \Big)^{-1} \big(\bm{A}_{pp}^\text{T} + i \omega \bm{I} \big)^{-1} \bm{A}_{ap}^\text{T} \bm{B}_a^{-1} \nn
&= \bm{B}_a^{-1} - \bm{B}_a^{-1} \bm{A}_{ap} \Big( \bm{A}_{ap}^\text{T} \bm{B}_a^{-1} \bm{A}_{ap} + \big(\bm{A}_{pp}^\text{T} + i \omega \bm{I} \big) \bm{B}_p^{-1} \big(\bm{A}_{pp} - i \omega \bm{I} \big) \Big)^{-1}  \bm{A}_{ap}^\text{T} \bm{B}_a^{-1}  . \n
\end{align}

\begin{figure*}
\includegraphics[width=0.4\textwidth]{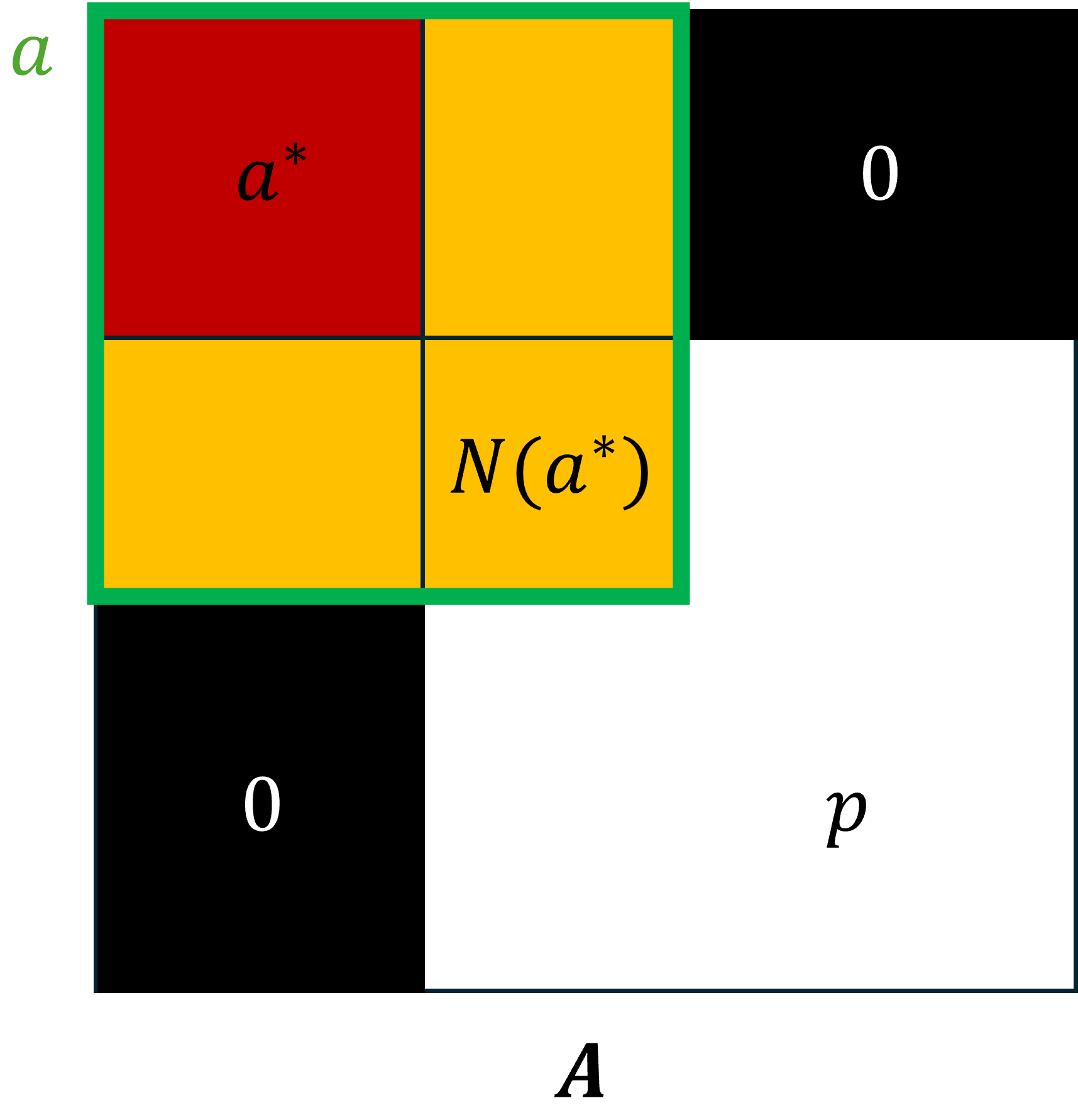}
\hspace{1cm}
\includegraphics[width=0.4\textwidth]{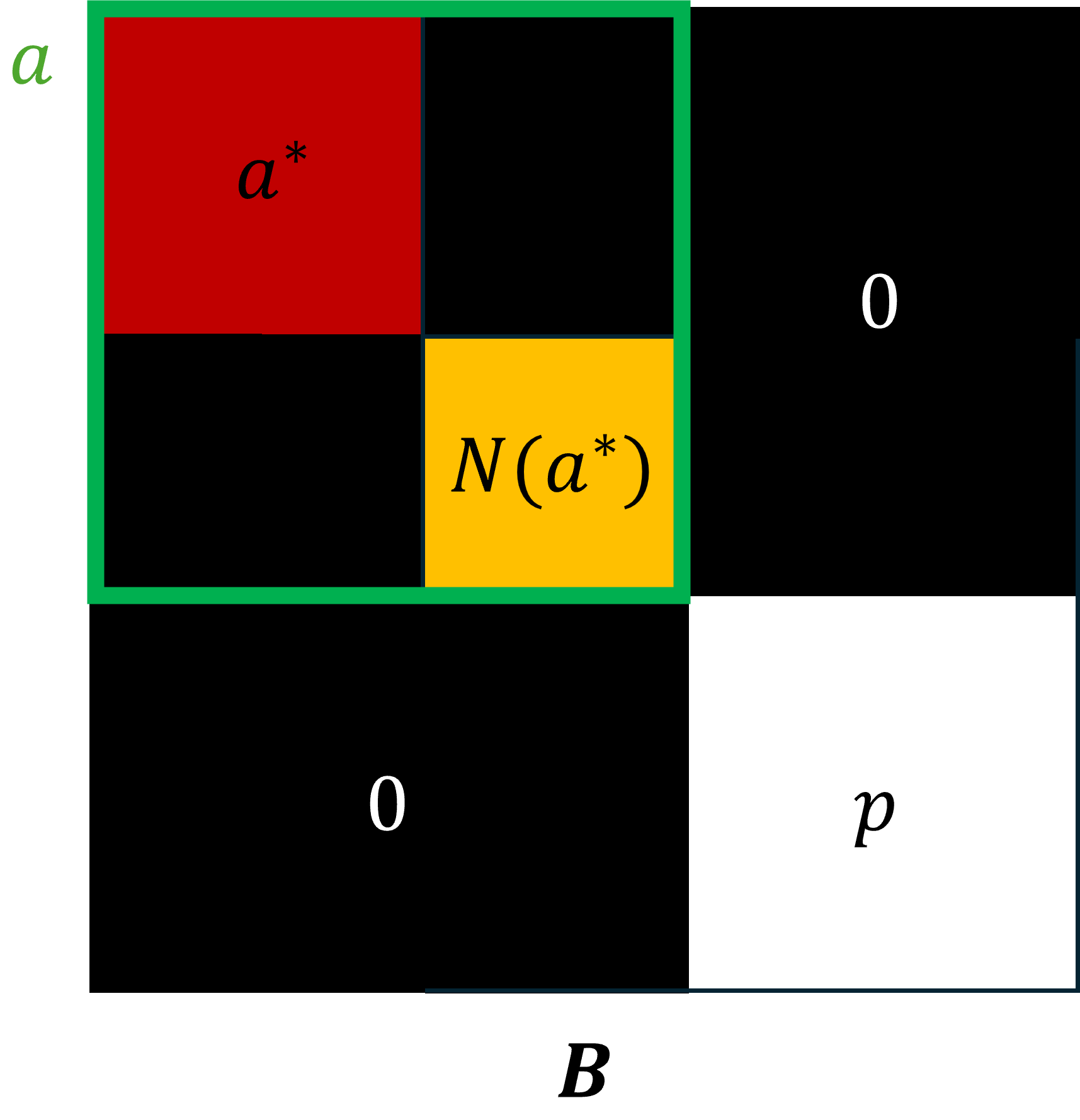}
\caption{The structure of the coupling matrix $\bm{A}$ (left) and diffusion matrix $\bm{B}$ (right). The perturbed degrees of freedom $a^*$ (red) and their neighbors $N(a^*)$ (yellow) are the measured degrees of freedom $a$ (green). Since all other degrees of freedom $p$ (white) are not connected to $a^*$, the corresponding elements of the coupling matrix $\bm{A}$ are zero. For the diffusion matrix, we assume a block-diagonal structure between the perturbed degrees of freedom $a^*$, their neighbors $N(a^*)$ and the remaining degrees of freedom $p$.}
\label{fig-partition}
\end{figure*}
Next, we define the connectivity of the system as follows:
We say that two degrees of freedom $x_j$ and $x_h$ are connected if $A_{jh} \neq 0$ or $A_{hj} \neq 0$, i.~e.~if $x_j$ and $x_h$ are not connected then $A_{jh} = A_{hj} = 0$.
We choose a set $a^* \subset \lbrace 1,\ldots,D \rbrace$ of degrees of freedom to which we apply the perturbation and define the set of neighbors $N(a^*)$ as all degrees of freedom that are connected to any degree of freedom in $a^*$ but not themselves contained in $a^*$, i.~e.
\begin{align}
N(a^*) = \lbrace j \vert h \in a^* \ \text{and} \ x_j \ \text{connected to} \ x_h \rbrace/a^* .
\end{align}
We now take $a = a^* \cup N(a^*)$, that is the perturbed degrees of freedom and their neighbors, as the \enquote{active} degrees of freedom.
In addition, we assume that the diffusion matrix $B_a$ is block-diagonal in $a^*$ and $N(a^*)$,
\begin{align}
\bm{B}_a = \begin{pmatrix} \bm{B}_{a^*} & \bm{0} \\ \bm{0} & \bm{B}_{N(a^*)} \end{pmatrix} ,
\end{align}
see Fig.~\ref{fig-partition} for an illustration.
Compared to the above, the difference is that we only apply the perturbation to a subset of degrees of freedom in $a$.
Since, by definition, $a$ includes all degrees of freedom connected to any degree of freedom in $a^*$, $\bm{A}_{a^*p} = \bm{A}_{p a^*} = 0$, allowing us to partition the coupling matrix as illustrated in Fig.~\ref{fig-partition}.
Since the perturbation is only acting on $a^*$, that is, only the rows of $\bmc{Y}$ corresponding to $\bm{a}^*$ are non-zero, which we indicate by writing $\bmc{Y}_{a^*}$, we find by combining \eqref{response-equality-4} and \eqref{woodbury},
\begin{align}
\bmc{R}_a(\omega)^\text{H} &\bmc{S}_{aa}(\omega)^{-1} \bmc{R}_a(\omega) = \frac{1}{2} \bmc{Y}_{a^*}^\text{T} \bigg( \Big( \bm{B}_a + \bm{A}_{ap} \big(\bm{A}_{pp} - i \omega \bm{I} \big)^{-1} \bm{B}_p \big(\bm{A}_{pp}^\text{T} + i \omega \bm{I} \big)^{-1} \bm{A}_{ap}^\text{T} \Big)^{-1} \bigg)_{a^* a^*} \bmc{Y}_{a^*} \\
&= \frac{1}{2} \bmc{Y}_{a^*}^\text{T} \Big( \bm{B}_{a^*}^{-1} - \bm{B}_{a^*}^{-1} \bm{A}_{a^*p} \Big( \bm{A}_{ap}^\text{T} \bm{B}_a^{-1} \bm{A}_{ap} + \big(\bm{A}_{pp}^\text{T} + i \omega \bm{I} \big) \bm{B}_p^{-1} \big(\bm{A}_{pp} - i \omega \bm{I} \big) \Big)^{-1}  \bm{A}_{a^*p}^\text{T} \bm{B}_{a^*}^{-1} \Big) \bmc{Y}_{a^*}  \nn
&= \frac{1}{2} \bmc{Y}_{a^*}^\text{T} \bm{B}_{a^*}^{-1} \bmc{Y}_{a^*} = \bmc{A} \n .
\end{align}
The above result implies that, under the assumption of a block-diagonal diffusion matrix (in particular for a diagonal diffusion matrix), it is sufficient to measure the perturbed degrees of freedom and their neighbors, i.~e.~degrees of freedom that directly interact with the perturbed ones, to realize equality in the finite-frequency FRI \eqref{app-fffri}.
This result can potentially be used to localize perturbations in linear networks:
If a measured set of degrees of freedom yields equality in \eqref{app-fffri}, this implies that the perturbation is localized in the interior of the set.
By dropping degrees of freedom from the measurement one by one and checking whether the equality is still satisfied, the degrees of freedom directly affected by the perturbation can be identified.

The above argument can also be applied to the underdamped case, since, by assumption, the matrices $\bm{m}$, $\bm{\gamma}$ and $\bm{T}$ are diagonal.
Thus, the only off-diagonal contributions stem from the coupling matrix $\bm{K}$, which has the same structure as the matrix $\bm{A}$ above. 
When computing the block-inverse to obtain the response function of the phase-space observables as in \eqref{response-block-inverse}, the only non-trivial contribution arises from the these non-diagonal parts of the coupling matrix.

\section{Response and fluctuations from Langevin simulations}
In the main text, we consider the diffusion of an underdamped particle in a tilted periodic potential, \eqref{langevin-perpot}.
We simulate this Langevin equation using the stochastic integrator developed in Ref.~\cite{Bussi2007}.
The velocity power-spectral density is calculated by computing the Fourier transform of the obtained time-discrete trajectories,
\begin{align}
\hat{v}_j = \Delta t \sum_{k = 1}^{N} e^{i \omega_j k \Delta t} v_k \quad \text{with} \quad \omega_j = \frac{2 \pi j}{\Delta t},
\end{align}
where $\Delta t$ is the time-step size and $v_k$, $k = 1,\ldots,N$, are the velocity values.
An estimate of the power-spectral density at $\omega_j$ is obtained as
\begin{align}
S_v(\omega_j) = \frac{1}{N \Delta t} \Av{\Vert \hat{v}_j \Vert^2},
\end{align}
where the average is taken over sample trajectories.
In order to account for aliasing due to the finite time-step \cite{Smith1999}, the power-spectral density shown in Fig.~\ref{fig-perpot} of the main text is the corrected spectral density,
\begin{align}
\bar{S}_v(\omega_j) = S_v(\omega_j) \frac{\sin\big(\frac{1}{2} \omega_j \Delta t\big)^2}{\big(\frac{1}{2} \omega_j \Delta t\big)^2 },
\end{align}
which corresponds to applying a lowpass filter to the signal.
In order to measure the frequency response, we apply a perturbation of the form $\epsilon g_0 \sin(\omega t)$ to the system.
We then fit the resulting average velocity with the function
\begin{align}
v_0 + \delta v \sin(\omega t + \phi) ,
\end{align}
where $v_0$, $\delta v$ and $\phi$ are fit parameters.
Since the real and imaginary parts of the response function correspond to the in-phase and out-of-phase response, respectively, they are related to the fit parameters as
\begin{align}
\Av{v}_\text{st} = v_0, \qquad \vert \bmc{R}_v(\omega) \vert = \frac{\delta v}{\epsilon}, \qquad \frac{\bmc{R}_v''(\omega)}{\bmc{R}_v'(\omega)} = \tan(\phi) ,
\end{align}
which, in particular, allows us to obtain the magnitude of the response $\vert \bmc{R}_v(\omega) \vert$ from the magnitude of the time-periodic part of the response.

\bibliography{../../../JabRef/JabRef_main.bib}

\begin{thebibliography}{53}%
\makeatletter
\providecommand \@ifxundefined [1]{%
 \@ifx{#1\undefined}
}%
\providecommand \@ifnum [1]{%
 \ifnum #1\expandafter \@firstoftwo
 \else \expandafter \@secondoftwo
 \fi
}%
\providecommand \@ifx [1]{%
 \ifx #1\expandafter \@firstoftwo
 \else \expandafter \@secondoftwo
 \fi
}%
\providecommand \natexlab [1]{#1}%
\providecommand \enquote  [1]{``#1''}%
\providecommand \bibnamefont  [1]{#1}%
\providecommand \bibfnamefont [1]{#1}%
\providecommand \citenamefont [1]{#1}%
\providecommand \href@noop [0]{\@secondoftwo}%
\providecommand \href [0]{\begingroup \@sanitize@url \@href}%
\providecommand \@href[1]{\@@startlink{#1}\@@href}%
\providecommand \@@href[1]{\endgroup#1\@@endlink}%
\providecommand \@sanitize@url [0]{\catcode `\\12\catcode `\$12\catcode
  `\&12\catcode `\#12\catcode `\^12\catcode `\_12\catcode `\%12\relax}%
\providecommand \@@startlink[1]{}%
\providecommand \@@endlink[0]{}%
\providecommand \url  [0]{\begingroup\@sanitize@url \@url }%
\providecommand \@url [1]{\endgroup\@href {#1}{\urlprefix }}%
\providecommand \urlprefix  [0]{URL }%
\providecommand \Eprint [0]{\href }%
\providecommand \doibase [0]{https://doi.org/}%
\providecommand \selectlanguage [0]{\@gobble}%
\providecommand \bibinfo  [0]{\@secondoftwo}%
\providecommand \bibfield  [0]{\@secondoftwo}%
\providecommand \translation [1]{[#1]}%
\providecommand \BibitemOpen [0]{}%
\providecommand \bibitemStop [0]{}%
\providecommand \bibitemNoStop [0]{.\EOS\space}%
\providecommand \EOS [0]{\spacefactor3000\relax}%
\providecommand \BibitemShut  [1]{\csname bibitem#1\endcsname}%
\let\auto@bib@innerbib\@empty
\bibitem [{\citenamefont {Kubo}(1966)}]{Kubo1966}%
  \BibitemOpen
  \bibfield  {author} {\bibinfo {author} {\bibfnamefont {R.}~\bibnamefont
  {Kubo}},\ }\bibfield  {title} {\bibinfo {title} {The fluctuation-dissipation
  theorem},\ }\href {http://stacks.iop.org/0034-4885/29/i=1/a=306} {\bibfield
  {journal} {\bibinfo  {journal} {Rep. Prog. Phys.}\ }\textbf {\bibinfo
  {volume} {29}},\ \bibinfo {pages} {255} (\bibinfo {year} {1966})}\BibitemShut
  {NoStop}%
\bibitem [{\citenamefont {Kubo}\ \emph {et~al.}(2012)\citenamefont {Kubo},
  \citenamefont {Toda},\ and\ \citenamefont {Hashitsume}}]{Kubo2012}%
  \BibitemOpen
  \bibfield  {author} {\bibinfo {author} {\bibfnamefont {R.}~\bibnamefont
  {Kubo}}, \bibinfo {author} {\bibfnamefont {M.}~\bibnamefont {Toda}},\ and\
  \bibinfo {author} {\bibfnamefont {N.}~\bibnamefont {Hashitsume}},\
  }\href@noop {} {\emph {\bibinfo {title} {Statistical {Physics} {II}:
  {Nonequilibrium} {Statistical} {Mechanics}}}}\ (\bibinfo  {publisher}
  {Springer Science \& Business Media},\ \bibinfo {year} {2012})\BibitemShut
  {NoStop}%
\bibitem [{\citenamefont {Cugliandolo}\ \emph {et~al.}(1997)\citenamefont
  {Cugliandolo}, \citenamefont {Kurchan},\ and\ \citenamefont
  {Peliti}}]{Cugliandolo1997}%
  \BibitemOpen
  \bibfield  {author} {\bibinfo {author} {\bibfnamefont {L.~F.}\ \bibnamefont
  {Cugliandolo}}, \bibinfo {author} {\bibfnamefont {J.}~\bibnamefont
  {Kurchan}},\ and\ \bibinfo {author} {\bibfnamefont {L.}~\bibnamefont
  {Peliti}},\ }\bibfield  {title} {\bibinfo {title} {Energy flow, partial
  equilibration, and effective temperatures in systems with slow dynamics},\
  }\href {https://doi.org/10.1103/PhysRevE.55.3898} {\bibfield  {journal}
  {\bibinfo  {journal} {Phys. Rev. E}\ }\textbf {\bibinfo {volume} {55}},\
  \bibinfo {pages} {3898} (\bibinfo {year} {1997})}\BibitemShut {NoStop}%
\bibitem [{\citenamefont {Loi}\ \emph {et~al.}(2008)\citenamefont {Loi},
  \citenamefont {Mossa},\ and\ \citenamefont {Cugliandolo}}]{Loi2008}%
  \BibitemOpen
  \bibfield  {author} {\bibinfo {author} {\bibfnamefont {D.}~\bibnamefont
  {Loi}}, \bibinfo {author} {\bibfnamefont {S.}~\bibnamefont {Mossa}},\ and\
  \bibinfo {author} {\bibfnamefont {L.~F.}\ \bibnamefont {Cugliandolo}},\
  }\bibfield  {title} {\bibinfo {title} {Effective temperature of active
  matter},\ }\href {https://doi.org/10.1103/PhysRevE.77.051111} {\bibfield
  {journal} {\bibinfo  {journal} {Physical Review E}\ }\textbf {\bibinfo
  {volume} {77}},\ \bibinfo {pages} {051111} (\bibinfo {year}
  {2008})}\BibitemShut {NoStop}%
\bibitem [{\citenamefont {Cugliandolo}(2011)}]{Cugliandolo2011}%
  \BibitemOpen
  \bibfield  {author} {\bibinfo {author} {\bibfnamefont {L.~F.}\ \bibnamefont
  {Cugliandolo}},\ }\bibfield  {title} {\bibinfo {title} {The effective
  temperature},\ }\href {https://doi.org/10.1088/1751-8113/44/48/483001}
  {\bibfield  {journal} {\bibinfo  {journal} {Journal of Physics A:
  Mathematical and Theoretical}\ }\textbf {\bibinfo {volume} {44}},\ \bibinfo
  {pages} {483001} (\bibinfo {year} {2011})}\BibitemShut {NoStop}%
\bibitem [{\citenamefont {Szamel}(2014)}]{Szamel2014}%
  \BibitemOpen
  \bibfield  {author} {\bibinfo {author} {\bibfnamefont {G.}~\bibnamefont
  {Szamel}},\ }\bibfield  {title} {\bibinfo {title} {Self-propelled particle in
  an external potential: {Existence} of an effective temperature},\ }\href
  {https://doi.org/10.1103/PhysRevE.90.012111} {\bibfield  {journal} {\bibinfo
  {journal} {Physical Review E}\ }\textbf {\bibinfo {volume} {90}},\ \bibinfo
  {pages} {012111} (\bibinfo {year} {2014})}\BibitemShut {NoStop}%
\bibitem [{\citenamefont {Lippiello}\ \emph {et~al.}(2014)\citenamefont
  {Lippiello}, \citenamefont {Baiesi},\ and\ \citenamefont
  {Sarracino}}]{Lippiello2014}%
  \BibitemOpen
  \bibfield  {author} {\bibinfo {author} {\bibfnamefont {E.}~\bibnamefont
  {Lippiello}}, \bibinfo {author} {\bibfnamefont {M.}~\bibnamefont {Baiesi}},\
  and\ \bibinfo {author} {\bibfnamefont {A.}~\bibnamefont {Sarracino}},\
  }\bibfield  {title} {\bibinfo {title} {Nonequilibrium
  {Fluctuation}-{Dissipation} {Theorem} and {Heat} {Production}},\ }\href
  {https://doi.org/10.1103/PhysRevLett.112.140602} {\bibfield  {journal}
  {\bibinfo  {journal} {Physical Review Letters}\ }\textbf {\bibinfo {volume}
  {112}},\ \bibinfo {pages} {140602} (\bibinfo {year} {2014})}\BibitemShut
  {NoStop}%
\bibitem [{\citenamefont {Fielding}\ and\ \citenamefont
  {Sollich}(2002)}]{Fielding2002}%
  \BibitemOpen
  \bibfield  {author} {\bibinfo {author} {\bibfnamefont {S.}~\bibnamefont
  {Fielding}}\ and\ \bibinfo {author} {\bibfnamefont {P.}~\bibnamefont
  {Sollich}},\ }\bibfield  {title} {\bibinfo {title} {Observable {Dependence}
  of {Fluctuation}-{Dissipation} {Relations} and {Effective} {Temperatures}},\
  }\href {https://doi.org/10.1103/PhysRevLett.88.050603} {\bibfield  {journal}
  {\bibinfo  {journal} {Physical Review Letters}\ }\textbf {\bibinfo {volume}
  {88}},\ \bibinfo {pages} {050603} (\bibinfo {year} {2002})}\BibitemShut
  {NoStop}%
\bibitem [{\citenamefont {Agarwal}(1972)}]{Agarwal1972}%
  \BibitemOpen
  \bibfield  {author} {\bibinfo {author} {\bibfnamefont {G.~S.}\ \bibnamefont
  {Agarwal}},\ }\bibfield  {title} {\bibinfo {title} {Fluctuation-dissipation
  theorems for systems in non-thermal equilibrium and applications},\ }\href
  {https://doi.org/10.1007/BF01391621} {\bibfield  {journal} {\bibinfo
  {journal} {Zeitschrift für Physik A Hadrons and nuclei}\ }\textbf {\bibinfo
  {volume} {252}},\ \bibinfo {pages} {25} (\bibinfo {year} {1972})}\BibitemShut
  {NoStop}%
\bibitem [{\citenamefont {Hänggi}\ and\ \citenamefont
  {Thomas}(1982)}]{Haenggi1982a}%
  \BibitemOpen
  \bibfield  {author} {\bibinfo {author} {\bibfnamefont {P.}~\bibnamefont
  {Hänggi}}\ and\ \bibinfo {author} {\bibfnamefont {H.}~\bibnamefont
  {Thomas}},\ }\bibfield  {title} {\bibinfo {title} {Stochastic processes:
  {Time} evolution, symmetries and linear response},\ }\href
  {https://doi.org/10.1016/0370-1573(82)90045-X} {\bibfield  {journal}
  {\bibinfo  {journal} {Physics Reports}\ }\textbf {\bibinfo {volume} {88}},\
  \bibinfo {pages} {207} (\bibinfo {year} {1982})}\BibitemShut {NoStop}%
\bibitem [{\citenamefont {Marconi}\ \emph {et~al.}(2008)\citenamefont
  {Marconi}, \citenamefont {Puglisi}, \citenamefont {Rondoni},\ and\
  \citenamefont {Vulpiani}}]{Marconi2008}%
  \BibitemOpen
  \bibfield  {author} {\bibinfo {author} {\bibfnamefont {U.~M.~B.}\
  \bibnamefont {Marconi}}, \bibinfo {author} {\bibfnamefont {A.}~\bibnamefont
  {Puglisi}}, \bibinfo {author} {\bibfnamefont {L.}~\bibnamefont {Rondoni}},\
  and\ \bibinfo {author} {\bibfnamefont {A.}~\bibnamefont {Vulpiani}},\
  }\bibfield  {title} {\bibinfo {title} {Fluctuation–dissipation: {Response}
  theory in statistical physics},\ }\href
  {https://doi.org/10.1016/j.physrep.2008.02.002} {\bibfield  {journal}
  {\bibinfo  {journal} {Physics Reports}\ }\textbf {\bibinfo {volume} {461}},\
  \bibinfo {pages} {111} (\bibinfo {year} {2008})}\BibitemShut {NoStop}%
\bibitem [{\citenamefont {Baiesi}\ \emph {et~al.}(2009)\citenamefont {Baiesi},
  \citenamefont {Maes},\ and\ \citenamefont {Wynants}}]{Baiesi2009}%
  \BibitemOpen
  \bibfield  {author} {\bibinfo {author} {\bibfnamefont {M.}~\bibnamefont
  {Baiesi}}, \bibinfo {author} {\bibfnamefont {C.}~\bibnamefont {Maes}},\ and\
  \bibinfo {author} {\bibfnamefont {B.}~\bibnamefont {Wynants}},\ }\bibfield
  {title} {\bibinfo {title} {Fluctuations and response of nonequilibrium
  states},\ }\href {https://doi.org/10.1103/PhysRevLett.103.010602} {\bibfield
  {journal} {\bibinfo  {journal} {Phys. Rev. Lett.}\ }\textbf {\bibinfo
  {volume} {103}},\ \bibinfo {pages} {010602} (\bibinfo {year}
  {2009})}\BibitemShut {NoStop}%
\bibitem [{\citenamefont {Seifert}\ and\ \citenamefont
  {Speck}(2010)}]{Seifert2010a}%
  \BibitemOpen
  \bibfield  {author} {\bibinfo {author} {\bibfnamefont {U.}~\bibnamefont
  {Seifert}}\ and\ \bibinfo {author} {\bibfnamefont {T.}~\bibnamefont
  {Speck}},\ }\bibfield  {title} {\bibinfo {title} {Fluctuation-dissipation
  theorem in nonequilibrium steady states},\ }\href
  {https://doi.org/10.1209/0295-5075/89/10007} {\bibfield  {journal} {\bibinfo
  {journal} {Europhysics Letters}\ }\textbf {\bibinfo {volume} {89}},\ \bibinfo
  {pages} {10007} (\bibinfo {year} {2010})}\BibitemShut {NoStop}%
\bibitem [{\citenamefont {Dechant}\ and\ \citenamefont
  {Sasa}(2020)}]{Dechant2020}%
  \BibitemOpen
  \bibfield  {author} {\bibinfo {author} {\bibfnamefont {A.}~\bibnamefont
  {Dechant}}\ and\ \bibinfo {author} {\bibfnamefont {S.-i.}\ \bibnamefont
  {Sasa}},\ }\bibfield  {title} {\bibinfo {title}
  {Fluctuation{\textendash}response inequality out of equilibrium},\ }\href
  {https://doi.org/10.1073/pnas.1918386117} {\bibfield  {journal} {\bibinfo
  {journal} {Proc. Natl. Acad. Sci.}\ }\textbf {\bibinfo {volume} {117}},\
  \bibinfo {pages} {6430} (\bibinfo {year} {2020})}\BibitemShut {NoStop}%
\bibitem [{\citenamefont {Owen}\ \emph {et~al.}(2020)\citenamefont {Owen},
  \citenamefont {Gingrich},\ and\ \citenamefont {Horowitz}}]{Owen2020}%
  \BibitemOpen
  \bibfield  {author} {\bibinfo {author} {\bibfnamefont {J.~A.}\ \bibnamefont
  {Owen}}, \bibinfo {author} {\bibfnamefont {T.~R.}\ \bibnamefont {Gingrich}},\
  and\ \bibinfo {author} {\bibfnamefont {J.~M.}\ \bibnamefont {Horowitz}},\
  }\bibfield  {title} {\bibinfo {title} {Universal {Thermodynamic} {Bounds} on
  {Nonequilibrium} {Response} with {Biochemical} {Applications}},\ }\href
  {https://doi.org/10.1103/PhysRevX.10.011066} {\bibfield  {journal} {\bibinfo
  {journal} {Physical Review X}\ }\textbf {\bibinfo {volume} {10}},\ \bibinfo
  {pages} {011066} (\bibinfo {year} {2020})}\BibitemShut {NoStop}%
\bibitem [{\citenamefont {Fernandes~Martins}\ and\ \citenamefont
  {Horowitz}(2023)}]{FernandesMartins2023}%
  \BibitemOpen
  \bibfield  {author} {\bibinfo {author} {\bibfnamefont {G.}~\bibnamefont
  {Fernandes~Martins}}\ and\ \bibinfo {author} {\bibfnamefont {J.~M.}\
  \bibnamefont {Horowitz}},\ }\bibfield  {title} {\bibinfo {title}
  {Topologically constrained fluctuations and thermodynamics regulate
  nonequilibrium response},\ }\href
  {https://doi.org/10.1103/PhysRevE.108.044113} {\bibfield  {journal} {\bibinfo
   {journal} {Physical Review E}\ }\textbf {\bibinfo {volume} {108}},\ \bibinfo
  {pages} {044113} (\bibinfo {year} {2023})}\BibitemShut {NoStop}%
\bibitem [{\citenamefont {Gao}\ \emph {et~al.}(2024)\citenamefont {Gao},
  \citenamefont {Chun},\ and\ \citenamefont {Horowitz}}]{Gao2024}%
  \BibitemOpen
  \bibfield  {author} {\bibinfo {author} {\bibfnamefont {Q.}~\bibnamefont
  {Gao}}, \bibinfo {author} {\bibfnamefont {H.-M.}\ \bibnamefont {Chun}},\ and\
  \bibinfo {author} {\bibfnamefont {J.~M.}\ \bibnamefont {Horowitz}},\
  }\bibfield  {title} {\bibinfo {title} {Thermodynamic constraints on kinetic
  perturbations of homogeneous driven diffusions},\ }\href
  {https://doi.org/10.1209/0295-5075/ad40cd} {\bibfield  {journal} {\bibinfo
  {journal} {Europhysics Letters}\ }\textbf {\bibinfo {volume} {146}},\
  \bibinfo {pages} {31001} (\bibinfo {year} {2024})}\BibitemShut {NoStop}%
\bibitem [{\citenamefont {Aslyamov}\ and\ \citenamefont
  {Esposito}(2024)}]{Aslyamov2024}%
  \BibitemOpen
  \bibfield  {author} {\bibinfo {author} {\bibfnamefont {T.}~\bibnamefont
  {Aslyamov}}\ and\ \bibinfo {author} {\bibfnamefont {M.}~\bibnamefont
  {Esposito}},\ }\bibfield  {title} {\bibinfo {title} {Nonequilibrium
  {Response} for {Markov} {Jump} {Processes}: {Exact} {Results} and {Tight}
  {Bounds}},\ }\href {https://doi.org/10.1103/PhysRevLett.132.037101}
  {\bibfield  {journal} {\bibinfo  {journal} {Physical Review Letters}\
  }\textbf {\bibinfo {volume} {132}},\ \bibinfo {pages} {037101} (\bibinfo
  {year} {2024})}\BibitemShut {NoStop}%
\bibitem [{\citenamefont {Ptaszyński}\ \emph {et~al.}(2024)\citenamefont
  {Ptaszyński}, \citenamefont {Aslyamov},\ and\ \citenamefont
  {Esposito}}]{Ptaszynski2024}%
  \BibitemOpen
  \bibfield  {author} {\bibinfo {author} {\bibfnamefont {K.}~\bibnamefont
  {Ptaszyński}}, \bibinfo {author} {\bibfnamefont {T.}~\bibnamefont
  {Aslyamov}},\ and\ \bibinfo {author} {\bibfnamefont {M.}~\bibnamefont
  {Esposito}},\ }\bibfield  {title} {\bibinfo {title} {Dissipation {Bounds}
  {Precision} of {Current} {Response} to {Kinetic} {Perturbations}},\ }\href
  {https://doi.org/10.1103/PhysRevLett.133.227101} {\bibfield  {journal}
  {\bibinfo  {journal} {Physical Review Letters}\ }\textbf {\bibinfo {volume}
  {133}},\ \bibinfo {pages} {227101} (\bibinfo {year} {2024})}\BibitemShut
  {NoStop}%
\bibitem [{\citenamefont {Liu}\ and\ \citenamefont {Gu}(2025)}]{Liu2025}%
  \BibitemOpen
  \bibfield  {author} {\bibinfo {author} {\bibfnamefont {K.}~\bibnamefont
  {Liu}}\ and\ \bibinfo {author} {\bibfnamefont {J.}~\bibnamefont {Gu}},\
  }\bibfield  {title} {\bibinfo {title} {Dynamical activity universally bounds
  precision of response in {Markovian} nonequilibrium systems},\ }\href
  {https://doi.org/10.1038/s42005-025-01982-w} {\bibfield  {journal} {\bibinfo
  {journal} {Communications Physics}\ }\textbf {\bibinfo {volume} {8}},\
  \bibinfo {pages} {62} (\bibinfo {year} {2025})}\BibitemShut {NoStop}%
\bibitem [{\citenamefont {Van~Vu}(2025)}]{VanVu2025}%
  \BibitemOpen
  \bibfield  {author} {\bibinfo {author} {\bibfnamefont {T.}~\bibnamefont
  {Van~Vu}},\ }\bibfield  {title} {\bibinfo {title} {Fundamental {Bounds} on
  {Precision} and {Response} for {Quantum} {Trajectory} {Observables}},\ }\href
  {https://doi.org/10.1103/PRXQuantum.6.010343} {\bibfield  {journal} {\bibinfo
   {journal} {PRX Quantum}\ }\textbf {\bibinfo {volume} {6}},\ \bibinfo {pages}
  {010343} (\bibinfo {year} {2025})}\BibitemShut {NoStop}%
\bibitem [{\citenamefont {Loi}\ \emph {et~al.}(2011)\citenamefont {Loi},
  \citenamefont {Mossa},\ and\ \citenamefont {Cugliandolo}}]{Loi2011a}%
  \BibitemOpen
  \bibfield  {author} {\bibinfo {author} {\bibfnamefont {D.}~\bibnamefont
  {Loi}}, \bibinfo {author} {\bibfnamefont {S.}~\bibnamefont {Mossa}},\ and\
  \bibinfo {author} {\bibfnamefont {L.~F.}\ \bibnamefont {Cugliandolo}},\
  }\bibfield  {title} {\bibinfo {title} {Effective temperature of active
  complex matter},\ }\href {https://doi.org/10.1039/C0SM01484B} {\bibfield
  {journal} {\bibinfo  {journal} {Soft Matter}\ }\textbf {\bibinfo {volume}
  {7}},\ \bibinfo {pages} {3726} (\bibinfo {year} {2011})}\BibitemShut
  {NoStop}%
\bibitem [{\citenamefont {Fodor}\ \emph {et~al.}(2016)\citenamefont {Fodor},
  \citenamefont {Nardini}, \citenamefont {Cates}, \citenamefont {Tailleur},
  \citenamefont {Visco},\ and\ \citenamefont {van Wijland}}]{Fodor2016}%
  \BibitemOpen
  \bibfield  {author} {\bibinfo {author} {\bibfnamefont {E.}~\bibnamefont
  {Fodor}}, \bibinfo {author} {\bibfnamefont {C.}~\bibnamefont {Nardini}},
  \bibinfo {author} {\bibfnamefont {M.~E.}\ \bibnamefont {Cates}}, \bibinfo
  {author} {\bibfnamefont {J.}~\bibnamefont {Tailleur}}, \bibinfo {author}
  {\bibfnamefont {P.}~\bibnamefont {Visco}},\ and\ \bibinfo {author}
  {\bibfnamefont {F.}~\bibnamefont {van Wijland}},\ }\bibfield  {title}
  {\bibinfo {title} {How {Far} from {Equilibrium} {Is} {Active} {Matter}?},\
  }\href {https://doi.org/10.1103/PhysRevLett.117.038103} {\bibfield  {journal}
  {\bibinfo  {journal} {Physical Review Letters}\ }\textbf {\bibinfo {volume}
  {117}},\ \bibinfo {pages} {038103} (\bibinfo {year} {2016})}\BibitemShut
  {NoStop}%
\bibitem [{\citenamefont {Song}\ \emph {et~al.}(2005)\citenamefont {Song},
  \citenamefont {Wang},\ and\ \citenamefont {Makse}}]{Song2005}%
  \BibitemOpen
  \bibfield  {author} {\bibinfo {author} {\bibfnamefont {C.}~\bibnamefont
  {Song}}, \bibinfo {author} {\bibfnamefont {P.}~\bibnamefont {Wang}},\ and\
  \bibinfo {author} {\bibfnamefont {H.~A.}\ \bibnamefont {Makse}},\ }\bibfield
  {title} {\bibinfo {title} {Experimental measurement of an effective
  temperature for jammed granular materials},\ }\href
  {https://doi.org/10.1073/pnas.0409911102} {\bibfield  {journal} {\bibinfo
  {journal} {Proceedings of the National Academy of Sciences}\ }\textbf
  {\bibinfo {volume} {102}},\ \bibinfo {pages} {2299} (\bibinfo {year}
  {2005})}\BibitemShut {NoStop}%
\bibitem [{\citenamefont {Sato}\ \emph {et~al.}(2003)\citenamefont {Sato},
  \citenamefont {Ito}, \citenamefont {Yomo},\ and\ \citenamefont
  {Kaneko}}]{Sato2003}%
  \BibitemOpen
  \bibfield  {author} {\bibinfo {author} {\bibfnamefont {K.}~\bibnamefont
  {Sato}}, \bibinfo {author} {\bibfnamefont {Y.}~\bibnamefont {Ito}}, \bibinfo
  {author} {\bibfnamefont {T.}~\bibnamefont {Yomo}},\ and\ \bibinfo {author}
  {\bibfnamefont {K.}~\bibnamefont {Kaneko}},\ }\bibfield  {title} {\bibinfo
  {title} {On the relation between fluctuation and response in biological
  systems},\ }\href {https://doi.org/10.1073/pnas.2334996100} {\bibfield
  {journal} {\bibinfo  {journal} {Proceedings of the National Academy of
  Sciences}\ }\textbf {\bibinfo {volume} {100}},\ \bibinfo {pages} {14086}
  (\bibinfo {year} {2003})}\BibitemShut {NoStop}%
\bibitem [{\citenamefont {Chen}\ \emph {et~al.}(2007)\citenamefont {Chen},
  \citenamefont {Lau}, \citenamefont {Hough}, \citenamefont {Islam},
  \citenamefont {Goulian}, \citenamefont {Lubensky},\ and\ \citenamefont
  {Yodh}}]{Chen2007}%
  \BibitemOpen
  \bibfield  {author} {\bibinfo {author} {\bibfnamefont {D.~T.~N.}\
  \bibnamefont {Chen}}, \bibinfo {author} {\bibfnamefont {A.~W.~C.}\
  \bibnamefont {Lau}}, \bibinfo {author} {\bibfnamefont {L.~A.}\ \bibnamefont
  {Hough}}, \bibinfo {author} {\bibfnamefont {M.~F.}\ \bibnamefont {Islam}},
  \bibinfo {author} {\bibfnamefont {M.}~\bibnamefont {Goulian}}, \bibinfo
  {author} {\bibfnamefont {T.~C.}\ \bibnamefont {Lubensky}},\ and\ \bibinfo
  {author} {\bibfnamefont {A.~G.}\ \bibnamefont {Yodh}},\ }\bibfield  {title}
  {\bibinfo {title} {Fluctuations and {Rheology} in {Active} {Bacterial}
  {Suspensions}},\ }\href {https://doi.org/10.1103/PhysRevLett.99.148302}
  {\bibfield  {journal} {\bibinfo  {journal} {Physical Review Letters}\
  }\textbf {\bibinfo {volume} {99}},\ \bibinfo {pages} {148302} (\bibinfo
  {year} {2007})}\BibitemShut {NoStop}%
\bibitem [{\citenamefont {Mizuno}\ \emph {et~al.}(2008)\citenamefont {Mizuno},
  \citenamefont {Head}, \citenamefont {MacKintosh},\ and\ \citenamefont
  {Schmidt}}]{Mizuno2008}%
  \BibitemOpen
  \bibfield  {author} {\bibinfo {author} {\bibfnamefont {D.}~\bibnamefont
  {Mizuno}}, \bibinfo {author} {\bibfnamefont {D.~A.}\ \bibnamefont {Head}},
  \bibinfo {author} {\bibfnamefont {F.~C.}\ \bibnamefont {MacKintosh}},\ and\
  \bibinfo {author} {\bibfnamefont {C.~F.}\ \bibnamefont {Schmidt}},\
  }\bibfield  {title} {\bibinfo {title} {Active and {Passive} {Microrheology}
  in {Equilibrium} and {Nonequilibrium} {Systems}},\ }\href
  {https://doi.org/10.1021/ma801218z} {\bibfield  {journal} {\bibinfo
  {journal} {Macromolecules}\ }\textbf {\bibinfo {volume} {41}},\ \bibinfo
  {pages} {7194} (\bibinfo {year} {2008})}\BibitemShut {NoStop}%
\bibitem [{\citenamefont {Hempston}\ \emph {et~al.}(2017)\citenamefont
  {Hempston}, \citenamefont {Vovrosh}, \citenamefont {Toroš}, \citenamefont
  {Winstone}, \citenamefont {Rashid},\ and\ \citenamefont
  {Ulbricht}}]{Hempston2017}%
  \BibitemOpen
  \bibfield  {author} {\bibinfo {author} {\bibfnamefont {D.}~\bibnamefont
  {Hempston}}, \bibinfo {author} {\bibfnamefont {J.}~\bibnamefont {Vovrosh}},
  \bibinfo {author} {\bibfnamefont {M.}~\bibnamefont {Toroš}}, \bibinfo
  {author} {\bibfnamefont {G.}~\bibnamefont {Winstone}}, \bibinfo {author}
  {\bibfnamefont {M.}~\bibnamefont {Rashid}},\ and\ \bibinfo {author}
  {\bibfnamefont {H.}~\bibnamefont {Ulbricht}},\ }\bibfield  {title} {\bibinfo
  {title} {Force sensing with an optically levitated charged nanoparticle},\
  }\href {https://doi.org/10.1063/1.4993555} {\bibfield  {journal} {\bibinfo
  {journal} {Applied Physics Letters}\ }\textbf {\bibinfo {volume} {111}},\
  \bibinfo {pages} {133111} (\bibinfo {year} {2017})}\BibitemShut {NoStop}%
\bibitem [{\citenamefont {Gonzalez-Ballestero}\ \emph
  {et~al.}(2021)\citenamefont {Gonzalez-Ballestero}, \citenamefont
  {Aspelmeyer}, \citenamefont {Novotny}, \citenamefont {Quidant},\ and\
  \citenamefont {Romero-Isart}}]{GonzalezBallestero2021}%
  \BibitemOpen
  \bibfield  {author} {\bibinfo {author} {\bibfnamefont {C.}~\bibnamefont
  {Gonzalez-Ballestero}}, \bibinfo {author} {\bibfnamefont {M.}~\bibnamefont
  {Aspelmeyer}}, \bibinfo {author} {\bibfnamefont {L.}~\bibnamefont {Novotny}},
  \bibinfo {author} {\bibfnamefont {R.}~\bibnamefont {Quidant}},\ and\ \bibinfo
  {author} {\bibfnamefont {O.}~\bibnamefont {Romero-Isart}},\ }\bibfield
  {title} {\bibinfo {title} {Levitodynamics: {Levitation} and control of
  microscopic objects in vacuum},\ }\href
  {https://doi.org/10.1126/science.abg3027} {\bibfield  {journal} {\bibinfo
  {journal} {Science}\ }\textbf {\bibinfo {volume} {374}},\ \bibinfo {pages}
  {eabg3027} (\bibinfo {year} {2021})}\BibitemShut {NoStop}%
\bibitem [{\citenamefont {Lau}\ \emph {et~al.}(2003)\citenamefont {Lau},
  \citenamefont {Hoffman}, \citenamefont {Davies}, \citenamefont {Crocker},\
  and\ \citenamefont {Lubensky}}]{Lau2003}%
  \BibitemOpen
  \bibfield  {author} {\bibinfo {author} {\bibfnamefont {A.~W.~C.}\
  \bibnamefont {Lau}}, \bibinfo {author} {\bibfnamefont {B.~D.}\ \bibnamefont
  {Hoffman}}, \bibinfo {author} {\bibfnamefont {A.}~\bibnamefont {Davies}},
  \bibinfo {author} {\bibfnamefont {J.~C.}\ \bibnamefont {Crocker}},\ and\
  \bibinfo {author} {\bibfnamefont {T.~C.}\ \bibnamefont {Lubensky}},\
  }\bibfield  {title} {\bibinfo {title} {Microrheology, {Stress}
  {Fluctuations}, and {Active} {Behavior} of {Living} {Cells}},\ }\href
  {https://doi.org/10.1103/PhysRevLett.91.198101} {\bibfield  {journal}
  {\bibinfo  {journal} {Physical Review Letters}\ }\textbf {\bibinfo {volume}
  {91}},\ \bibinfo {pages} {198101} (\bibinfo {year} {2003})}\BibitemShut
  {NoStop}%
\bibitem [{\citenamefont {Brau}\ \emph {et~al.}(2007)\citenamefont {Brau},
  \citenamefont {Ferrer}, \citenamefont {Lee}, \citenamefont {Castro},
  \citenamefont {Tam}, \citenamefont {Tarsa}, \citenamefont {Matsudaira},
  \citenamefont {Boyce}, \citenamefont {Kamm},\ and\ \citenamefont
  {Lang}}]{Brau2007}%
  \BibitemOpen
  \bibfield  {author} {\bibinfo {author} {\bibfnamefont {R.~R.}\ \bibnamefont
  {Brau}}, \bibinfo {author} {\bibfnamefont {J.~M.}\ \bibnamefont {Ferrer}},
  \bibinfo {author} {\bibfnamefont {H.}~\bibnamefont {Lee}}, \bibinfo {author}
  {\bibfnamefont {C.~E.}\ \bibnamefont {Castro}}, \bibinfo {author}
  {\bibfnamefont {B.~K.}\ \bibnamefont {Tam}}, \bibinfo {author} {\bibfnamefont
  {P.~B.}\ \bibnamefont {Tarsa}}, \bibinfo {author} {\bibfnamefont
  {P.}~\bibnamefont {Matsudaira}}, \bibinfo {author} {\bibfnamefont {M.~C.}\
  \bibnamefont {Boyce}}, \bibinfo {author} {\bibfnamefont {R.~D.}\ \bibnamefont
  {Kamm}},\ and\ \bibinfo {author} {\bibfnamefont {M.~J.}\ \bibnamefont
  {Lang}},\ }\bibfield  {title} {\bibinfo {title} {Passive and active
  microrheology with optical tweezers},\ }\href
  {https://doi.org/10.1088/1464-4258/9/8/S01} {\bibfield  {journal} {\bibinfo
  {journal} {Journal of Optics A: Pure and Applied Optics}\ }\textbf {\bibinfo
  {volume} {9}},\ \bibinfo {pages} {S103} (\bibinfo {year} {2007})}\BibitemShut
  {NoStop}%
\bibitem [{\citenamefont {Wilhelm}(2008)}]{Wilhelm2008}%
  \BibitemOpen
  \bibfield  {author} {\bibinfo {author} {\bibfnamefont {C.}~\bibnamefont
  {Wilhelm}},\ }\bibfield  {title} {\bibinfo {title} {Out-of-{Equilibrium}
  {Microrheology} inside {Living} {Cells}},\ }\href
  {https://doi.org/10.1103/PhysRevLett.101.028101} {\bibfield  {journal}
  {\bibinfo  {journal} {Physical Review Letters}\ }\textbf {\bibinfo {volume}
  {101}},\ \bibinfo {pages} {028101} (\bibinfo {year} {2008})}\BibitemShut
  {NoStop}%
\bibitem [{\citenamefont {Ben-Isaac}\ \emph {et~al.}(2011)\citenamefont
  {Ben-Isaac}, \citenamefont {Park}, \citenamefont {Popescu}, \citenamefont
  {Brown}, \citenamefont {Gov},\ and\ \citenamefont {Shokef}}]{BenIsaac2011}%
  \BibitemOpen
  \bibfield  {author} {\bibinfo {author} {\bibfnamefont {E.}~\bibnamefont
  {Ben-Isaac}}, \bibinfo {author} {\bibfnamefont {Y.}~\bibnamefont {Park}},
  \bibinfo {author} {\bibfnamefont {G.}~\bibnamefont {Popescu}}, \bibinfo
  {author} {\bibfnamefont {F.~L.~H.}\ \bibnamefont {Brown}}, \bibinfo {author}
  {\bibfnamefont {N.~S.}\ \bibnamefont {Gov}},\ and\ \bibinfo {author}
  {\bibfnamefont {Y.}~\bibnamefont {Shokef}},\ }\bibfield  {title} {\bibinfo
  {title} {Effective {Temperature} of {Red}-{Blood}-{Cell} {Membrane}
  {Fluctuations}},\ }\href {https://doi.org/10.1103/PhysRevLett.106.238103}
  {\bibfield  {journal} {\bibinfo  {journal} {Physical Review Letters}\
  }\textbf {\bibinfo {volume} {106}},\ \bibinfo {pages} {238103} (\bibinfo
  {year} {2011})}\BibitemShut {NoStop}%
\bibitem [{\citenamefont {Risken}(1986)}]{Risken1986}%
  \BibitemOpen
  \bibfield  {author} {\bibinfo {author} {\bibfnamefont {H.}~\bibnamefont
  {Risken}},\ }\href@noop {} {\emph {\bibinfo {title} {The Fokker-Planck
  Equation}}}\ (\bibinfo  {publisher} {Springer Berlin},\ \bibinfo {year}
  {1986})\BibitemShut {NoStop}%
\bibitem [{\citenamefont {Coffey}\ and\ \citenamefont
  {Kalmykov}(2017)}]{Coffey2017}%
  \BibitemOpen
  \bibfield  {author} {\bibinfo {author} {\bibfnamefont {W.~M.}\ \bibnamefont
  {Coffey}}\ and\ \bibinfo {author} {\bibfnamefont {Y.~P.}\ \bibnamefont
  {Kalmykov}},\ }\href@noop {} {\emph {\bibinfo {title} {The Langevin equation:
  with applications to stochastic problems in physics, chemistry, and
  electrical engineering}}}\ (\bibinfo  {publisher} {World Scientific},\
  \bibinfo {year} {2017})\BibitemShut {NoStop}%
\bibitem [{\citenamefont {Stratonovich}(1963)}]{Stratonovich1963}%
  \BibitemOpen
  \bibfield  {author} {\bibinfo {author} {\bibfnamefont {R.~L.}\ \bibnamefont
  {Stratonovich}},\ }\href@noop {} {\emph {\bibinfo {title} {Topics in the
  {Theory} of {Random} {Noise}}}}\ (\bibinfo  {publisher} {Gordon \& Breach
  Publishing Group},\ \bibinfo {year} {1963})\BibitemShut {NoStop}%
\bibitem [{\citenamefont {Vollmer}\ and\ \citenamefont
  {Risken}(1983)}]{Vollmer1983}%
  \BibitemOpen
  \bibfield  {author} {\bibinfo {author} {\bibfnamefont {H.~D.}\ \bibnamefont
  {Vollmer}}\ and\ \bibinfo {author} {\bibfnamefont {H.}~\bibnamefont
  {Risken}},\ }\bibfield  {title} {\bibinfo {title} {Eigenvalues and their
  connection to transition rates for the {Brownian} motion in an inclined
  cosine potential},\ }\href {https://doi.org/10.1007/BF01307378} {\bibfield
  {journal} {\bibinfo  {journal} {Zeitschrift für Physik B Condensed Matter}\
  }\textbf {\bibinfo {volume} {52}},\ \bibinfo {pages} {259} (\bibinfo {year}
  {1983})}\BibitemShut {NoStop}%
\bibitem [{\citenamefont {Lindenberg}\ \emph {et~al.}(2005)\citenamefont
  {Lindenberg}, \citenamefont {Lacasta}, \citenamefont {Sancho},\ and\
  \citenamefont {Romero}}]{Lindenberg2005}%
  \BibitemOpen
  \bibfield  {author} {\bibinfo {author} {\bibfnamefont {K.}~\bibnamefont
  {Lindenberg}}, \bibinfo {author} {\bibfnamefont {A.~M.}\ \bibnamefont
  {Lacasta}}, \bibinfo {author} {\bibfnamefont {J.~M.}\ \bibnamefont
  {Sancho}},\ and\ \bibinfo {author} {\bibfnamefont {A.~H.}\ \bibnamefont
  {Romero}},\ }\bibfield  {title} {\bibinfo {title} {Transport and diffusion on
  crystalline surfaces under external forces},\ }\href
  {https://doi.org/10.1088/1367-2630/7/1/029} {\bibfield  {journal} {\bibinfo
  {journal} {New Journal of Physics}\ }\textbf {\bibinfo {volume} {7}},\
  \bibinfo {pages} {29} (\bibinfo {year} {2005})}\BibitemShut {NoStop}%
\bibitem [{\citenamefont {Marchenko}\ and\ \citenamefont
  {Marchenko}(2012)}]{Marchenko2012}%
  \BibitemOpen
  \bibfield  {author} {\bibinfo {author} {\bibfnamefont {I.~G.}\ \bibnamefont
  {Marchenko}}\ and\ \bibinfo {author} {\bibfnamefont {I.~I.}\ \bibnamefont
  {Marchenko}},\ }\bibfield  {title} {\bibinfo {title} {Diffusion in the
  systems with low dissipation: {Exponential} growth with temperature drop},\
  }\href {https://doi.org/10.1209/0295-5075/100/50005} {\bibfield  {journal}
  {\bibinfo  {journal} {Europhysics Letters}\ }\textbf {\bibinfo {volume}
  {100}},\ \bibinfo {pages} {50005} (\bibinfo {year} {2012})}\BibitemShut
  {NoStop}%
\bibitem [{\citenamefont {Lindner}\ and\ \citenamefont
  {Sokolov}(2016)}]{Lindner2016}%
  \BibitemOpen
  \bibfield  {author} {\bibinfo {author} {\bibfnamefont {B.}~\bibnamefont
  {Lindner}}\ and\ \bibinfo {author} {\bibfnamefont {I.~M.}\ \bibnamefont
  {Sokolov}},\ }\bibfield  {title} {\bibinfo {title} {Giant diffusion of
  underdamped particles in a biased periodic potential},\ }\href
  {https://doi.org/10.1103/PhysRevE.93.042106} {\bibfield  {journal} {\bibinfo
  {journal} {Physical Review E}\ }\textbf {\bibinfo {volume} {93}},\ \bibinfo
  {pages} {042106} (\bibinfo {year} {2016})}\BibitemShut {NoStop}%
\bibitem [{\citenamefont {Fischer}\ \emph {et~al.}(2018)\citenamefont
  {Fischer}, \citenamefont {Pietzonka},\ and\ \citenamefont
  {Seifert}}]{Fischer2018}%
  \BibitemOpen
  \bibfield  {author} {\bibinfo {author} {\bibfnamefont {L.~P.}\ \bibnamefont
  {Fischer}}, \bibinfo {author} {\bibfnamefont {P.}~\bibnamefont {Pietzonka}},\
  and\ \bibinfo {author} {\bibfnamefont {U.}~\bibnamefont {Seifert}},\
  }\bibfield  {title} {\bibinfo {title} {Large deviation function for a driven
  underdamped particle in a periodic potential},\ }\href
  {https://doi.org/10.1103/PhysRevE.97.022143} {\bibfield  {journal} {\bibinfo
  {journal} {Physical Review E}\ }\textbf {\bibinfo {volume} {97}},\ \bibinfo
  {pages} {022143} (\bibinfo {year} {2018})}\BibitemShut {NoStop}%
\bibitem [{\citenamefont {Fischer}\ \emph {et~al.}(2020)\citenamefont
  {Fischer}, \citenamefont {Chun},\ and\ \citenamefont
  {Seifert}}]{Fischer2020}%
  \BibitemOpen
  \bibfield  {author} {\bibinfo {author} {\bibfnamefont {L.~P.}\ \bibnamefont
  {Fischer}}, \bibinfo {author} {\bibfnamefont {H.-M.}\ \bibnamefont {Chun}},\
  and\ \bibinfo {author} {\bibfnamefont {U.}~\bibnamefont {Seifert}},\
  }\bibfield  {title} {\bibinfo {title} {Free diffusion bounds the precision of
  currents in underdamped dynamics},\ }\href
  {https://doi.org/10.1103/PhysRevE.102.012120} {\bibfield  {journal} {\bibinfo
   {journal} {Physical Review E}\ }\textbf {\bibinfo {volume} {102}},\ \bibinfo
  {pages} {012120} (\bibinfo {year} {2020})}\BibitemShut {NoStop}%
\bibitem [{\citenamefont {Spiechowicz}\ \emph {et~al.}(2023)\citenamefont
  {Spiechowicz}, \citenamefont {Marchenko}, \citenamefont {Hänggi},\ and\
  \citenamefont {Łuczka}}]{Spiechowicz2023}%
  \BibitemOpen
  \bibfield  {author} {\bibinfo {author} {\bibfnamefont {J.}~\bibnamefont
  {Spiechowicz}}, \bibinfo {author} {\bibfnamefont {I.~G.}\ \bibnamefont
  {Marchenko}}, \bibinfo {author} {\bibfnamefont {P.}~\bibnamefont {Hänggi}},\
  and\ \bibinfo {author} {\bibfnamefont {J.}~\bibnamefont {Łuczka}},\
  }\bibfield  {title} {\bibinfo {title} {Diffusion {Coefficient} of a
  {Brownian} {Particle} in {Equilibrium} and {Nonequilibrium}: {Einstein}
  {Model} and {Beyond}},\ }\href {https://doi.org/10.3390/e25010042} {\bibfield
   {journal} {\bibinfo  {journal} {Entropy}\ }\textbf {\bibinfo {volume}
  {25}},\ \bibinfo {pages} {42} (\bibinfo {year} {2023})}\BibitemShut {NoStop}%
\bibitem [{\citenamefont {Reimann}\ \emph {et~al.}(2001)\citenamefont
  {Reimann}, \citenamefont {Van~den Broeck}, \citenamefont {Linke},
  \citenamefont {H\"anggi}, \citenamefont {Rubi},\ and\ \citenamefont
  {P\'erez-Madrid}}]{Reimann2001}%
  \BibitemOpen
  \bibfield  {author} {\bibinfo {author} {\bibfnamefont {P.}~\bibnamefont
  {Reimann}}, \bibinfo {author} {\bibfnamefont {C.}~\bibnamefont {Van~den
  Broeck}}, \bibinfo {author} {\bibfnamefont {H.}~\bibnamefont {Linke}},
  \bibinfo {author} {\bibfnamefont {P.}~\bibnamefont {H\"anggi}}, \bibinfo
  {author} {\bibfnamefont {J.~M.}\ \bibnamefont {Rubi}},\ and\ \bibinfo
  {author} {\bibfnamefont {A.}~\bibnamefont {P\'erez-Madrid}},\ }\bibfield
  {title} {\bibinfo {title} {Giant acceleration of free diffusion by use of
  tilted periodic potentials},\ }\href
  {https://doi.org/10.1103/PhysRevLett.87.010602} {\bibfield  {journal}
  {\bibinfo  {journal} {Phys. Rev. Lett.}\ }\textbf {\bibinfo {volume} {87}},\
  \bibinfo {pages} {010602} (\bibinfo {year} {2001})}\BibitemShut {NoStop}%
\bibitem [{\citenamefont {Fraden}(2010)}]{Fraden2010}%
  \BibitemOpen
  \bibfield  {author} {\bibinfo {author} {\bibfnamefont {J.}~\bibnamefont
  {Fraden}},\ }\href@noop {} {\emph {\bibinfo {title} {Handbook of {Modern}
  {Sensors}: {Physics}, {Designs}, and {Applications}}}}\ (\bibinfo
  {publisher} {Springer Science \& Business Media},\ \bibinfo {year} {2010})\
  \bibinfo {note} {google-Books-ID: W0Emv9dAJ1kC}\BibitemShut {NoStop}%
\bibitem [{\citenamefont {Dechant}\ and\ \citenamefont
  {Lutz}(2025)}]{Dechant2024}%
  \BibitemOpen
  \bibfield  {author} {\bibinfo {author} {\bibfnamefont {A.}~\bibnamefont
  {Dechant}}\ and\ \bibinfo {author} {\bibfnamefont {E.}~\bibnamefont {Lutz}},\
  }\bibfield  {title} {\bibinfo {title} {Fundamental limits on nonequilibrium
  sensing},\ }\bibfield  {journal} {\bibinfo  {journal} {Nature Communications
  (accepted)}\ }\href {https://doi.org/10.48550/arXiv.2407.17831}
  {10.48550/arXiv.2407.17831} (\bibinfo {year} {2025})\BibitemShut {NoStop}%
\bibitem [{\citenamefont {Harada}\ and\ \citenamefont
  {Sasa}(2005)}]{Harada2005}%
  \BibitemOpen
  \bibfield  {author} {\bibinfo {author} {\bibfnamefont {T.}~\bibnamefont
  {Harada}}\ and\ \bibinfo {author} {\bibfnamefont {S.-i.}\ \bibnamefont
  {Sasa}},\ }\bibfield  {title} {\bibinfo {title} {Equality connecting energy
  dissipation with a violation of the fluctuation-response relation},\ }\href
  {https://doi.org/10.1103/PhysRevLett.95.130602} {\bibfield  {journal}
  {\bibinfo  {journal} {Physical Review Letters}\ }\textbf {\bibinfo {volume}
  {95}},\ \bibinfo {pages} {130602} (\bibinfo {year} {2005})}\BibitemShut
  {NoStop}%
\bibitem [{\citenamefont {Harada}\ and\ \citenamefont
  {Sasa}(2006)}]{Harada2006}%
  \BibitemOpen
  \bibfield  {author} {\bibinfo {author} {\bibfnamefont {T.}~\bibnamefont
  {Harada}}\ and\ \bibinfo {author} {\bibfnamefont {S.-i.}\ \bibnamefont
  {Sasa}},\ }\bibfield  {title} {\bibinfo {title} {Energy dissipation and
  violation of the fluctuation-response relation in nonequilibrium {Langevin}
  systems},\ }\href {https://doi.org/10.1103/PhysRevE.73.026131} {\bibfield
  {journal} {\bibinfo  {journal} {Physical Review E}\ }\textbf {\bibinfo
  {volume} {73}},\ \bibinfo {pages} {026131} (\bibinfo {year}
  {2006})}\BibitemShut {NoStop}%
\bibitem [{\citenamefont {Bao}\ and\ \citenamefont {Liang}(2025)}]{Bao2025}%
  \BibitemOpen
  \bibfield  {author} {\bibinfo {author} {\bibfnamefont {R.}~\bibnamefont
  {Bao}}\ and\ \bibinfo {author} {\bibfnamefont {S.}~\bibnamefont {Liang}},\
  }\href {https://doi.org/10.48550/arXiv.2412.19602} {\bibinfo {title}
  {Nonlinear {Response} {Identities} and {Bounds} for {Nonequilibrium} {Steady}
  {States}}} (\bibinfo {year} {2025}),\ \bibinfo {note} {arXiv:2412.19602
  [cond-mat]}\BibitemShut {NoStop}%
\bibitem [{\citenamefont {Dechant}(2023)}]{Dechant2023a}%
  \BibitemOpen
  \bibfield  {author} {\bibinfo {author} {\bibfnamefont {A.}~\bibnamefont
  {Dechant}},\ }\bibfield  {title} {\bibinfo {title} {Thermodynamic constraints
  on the power spectral density in and out of equilibrium},\ }\href@noop {}
  {\bibfield  {journal} {\bibinfo  {journal} {arXiv preprint arXiv:2306.00417}\
  } (\bibinfo {year} {2023})}\BibitemShut {NoStop}%
\bibitem [{\citenamefont {Vo}\ \emph {et~al.}(2025)\citenamefont {Vo},
  \citenamefont {Dechant},\ and\ \citenamefont {Saito}}]{Vo2025}%
  \BibitemOpen
  \bibfield  {author} {\bibinfo {author} {\bibfnamefont {V.~T.}\ \bibnamefont
  {Vo}}, \bibinfo {author} {\bibfnamefont {A.}~\bibnamefont {Dechant}},\ and\
  \bibinfo {author} {\bibfnamefont {K.}~\bibnamefont {Saito}},\ }\href
  {https://doi.org/10.48550/arXiv.2503.14204} {\bibinfo {title} {Inverse
  thermodynamic uncertainty relation and entropy production}} (\bibinfo {year}
  {2025}),\ \bibinfo {note} {arXiv:2503.14204 [cond-mat]}\BibitemShut {NoStop}%
\bibitem [{\citenamefont {Bussi}\ and\ \citenamefont
  {Parrinello}(2007)}]{Bussi2007}%
  \BibitemOpen
  \bibfield  {author} {\bibinfo {author} {\bibfnamefont {G.}~\bibnamefont
  {Bussi}}\ and\ \bibinfo {author} {\bibfnamefont {M.}~\bibnamefont
  {Parrinello}},\ }\bibfield  {title} {\bibinfo {title} {Accurate sampling
  using {Langevin} dynamics},\ }\href
  {https://doi.org/10.1103/PhysRevE.75.056707} {\bibfield  {journal} {\bibinfo
  {journal} {Physical Review E}\ }\textbf {\bibinfo {volume} {75}},\ \bibinfo
  {pages} {056707} (\bibinfo {year} {2007})}\BibitemShut {NoStop}%
\bibitem [{\citenamefont {Smith}(1999)}]{Smith1999}%
  \BibitemOpen
  \bibfield  {author} {\bibinfo {author} {\bibfnamefont {S.~W.}\ \bibnamefont
  {Smith}},\ }\href@noop {} {\emph {\bibinfo {title} {The Scientist and
  Engineer’s Guide to Digital Signal Processing}}}\ (\bibinfo  {publisher}
  {California Technical Publishing},\ \bibinfo {year} {1999})\BibitemShut
  {NoStop}%
\end{thebibliography}%

\end{document}